\documentclass[11pt]{article}
\usepackage[margin=1in]{geometry}
\usepackage[T1]{fontenc}
\usepackage{lmodern,microtype}
\usepackage{amsmath,amssymb,amsthm,mathtools}
\usepackage{enumitem,xcolor,etoolbox}
\usepackage[colorlinks=true,linkcolor=blue!45!black,citecolor=blue!45!black,urlcolor=blue!45!black]{hyperref}
\hypersetup{pdftitle={Continual Differential Privacy versus Retroactivity}}
\setlist{itemsep=3pt,topsep=5pt}
\newtheorem{theorem}{Theorem}[section]
\newtheorem{lemma}[theorem]{Lemma}
\newtheorem{observation}[theorem]{Observation}

\newtheorem{corollary}[theorem]{Corollary}
\newtheorem{definition}[theorem]{Definition}
\newtheorem{remark}[theorem]{Remark}
\newcommand{\E}{\mathbb E}
\newcommand{\Prb}{\mathbb P}

\newcommand{\eps}{\varepsilon}
\newcommand{\one}{\mathbf 1}
\newcommand{\clip}{\operatorname{clip}}
\newcommand{\Lap}{\operatorname{Lap}}

\DeclareMathOperator{\Law}{Law}
\patchcmd{\thebibliography}{\section*{\refname}}{\paragraph{References}}{}{\PackageError{paper}{Bibliography heading patch failed}{}}
\DeclareMathOperator{\sgn}{sgn}

\title{Continual Differential Privacy versus Retroactivity}
\author{
Edith Cohen\textsuperscript{$\ddagger,*$}
\and
Travis Dick\textsuperscript{$\ddagger$}
\and
Haim Kaplan\textsuperscript{$*,\ddagger$}
\and
Refael Kohen\textsuperscript{$*$}
\and
Alex Kulesza\textsuperscript{$\ddagger$}
\and
Kobbi Nissim\textsuperscript{$\dagger, \ddagger$}
\and
Uri Stemmer\textsuperscript{$*,\ddagger$}
}
\date{}

\begin{document}
\maketitle

\begingroup
\renewcommand{\thefootnote}{} 
\footnotetext{
  \hspace{-7px}
  \textsuperscript{*}Tel Aviv University \quad
  \textsuperscript{$\dagger$}Georgetown University \quad
  \textsuperscript{$\ddagger$}Google Research 
}
\endgroup

\begin{abstract}
We give a separation between DP under
continual observation and the additional requirement of
{\em retroactivity}, a formalization of machine unlearning requiring
future outputs to have approximately the same joint distribution
as if deleted records had never been inserted. Specifically, we present a scalar function that, for a time horizon $T$, can be approximated under continual DP with simultaneous error $O({\rm polylog}(T))$, but every continually DP and retroactive algorithm requires error $\Omega({\rm poly}(T))$. This resolves an open question of Kaplan et al.~(2026).

Our lower bound has implications for the {\em privacy for the analyst} model, where the goal is to answer online queries while providing privacy both for the data and for the queries of other analysts, partially resolving an open question of Dwork et al.~(2012).
\end{abstract}

\section{Introduction}

Motivated by privacy regulations such as the GDPR, which recognizes the ``right to be forgotten'', the line of work on {\em machine unlearning} aims to allow users to remove their data from the training set of an already trained model. Cao and Yang~\cite{CY15} defined such a removal as successful if the resulting model is distributed identically to a model retrained from scratch without the removed data. However, as was shown by several recent works, this requirement alone does not guarantee privacy against an observer who sees several releases over time~\cite{CZW+21,CKNS26}. Indeed, comparing releases before and after a deletion can enable ``differencing attacks'' that reveal information about the deleted record. Moreover, and perhaps more surprisingly, such deletion requests may also reveal a lot of information about the {\em records that remain} in the dataset (the ``undeleted'' records).

Thus, arguably, this notion of machine unlearning is better understood as a consistency requirement than as a privacy requirement. In fact, as observed by Kaplan et al.~\cite{KKMNS26}, this underlying consistency requirement predates the literature on machine unlearning and was studied by the dynamic algorithms community under the term {\em retroactive data structures}~\cite{DIL07}. These are data structures supporting modifications to past operations while requiring subsequent answers to reflect the revised history as if it had always been in force.

This motivated Kaplan et al.~\cite{KKMNS26} to introduce the notion of {\em private retroactive algorithms}, combining retroactivity with differential privacy (DP) under continual observation. The privacy constraint requires the joint distribution of the entire output sequence to be insensitive to removing a single update. The retroactivity constraint requires the joint distribution of future answers to be the same as if all previously deleted records had never been inserted, regardless of how many such deletions occurred.

Kaplan et al.~\cite{KKMNS26} showed that, in many cases, the price of imposing both retroactivity and continual-DP is essentially bounded by the price of imposing continual-DP alone. In addition, they showed a generic construction that transforms a continually-DP algorithm into one that also satisfies \emph{marginal approximate retroactivity} with very little overhead in the error. In the former cases, therefore, retroactivity comes ``for free'' compared to the costs of continual-DP alone. Kaplan et al.~\cite{KKMNS26} asked whether retroactivity always comes ``for free'' in terms of worst-case error.

We answer negatively by constructing a scalar function that, over a time horizon $T$, admits a continually-DP algorithm with simultaneous error $O({\rm polylog}(T))$, but for which every continually-DP and retroactive algorithm requires worst-case error $\Omega({\rm poly}(T))$.

\subsection{The Model}
\label{sec:model}
Before describing our new results, we make our setting explicit. 
Fix a public horizon $T$ and an initially empty database. At each time
there is one insertion, one deletion request, or a no-op.
A deletion request specifies a record $r$ and the time $\tau$ at which
that occurrence was inserted into the stream. It removes that occurrence
if it is still present; otherwise, the request has no effect.
Let $D_t(\mathbf u)$ be the multiset of records present at time $t$
in a stream $\mathbf u$, each paired with its insertion time.
Neighboring streams differ by replacing one operation
with a no-op, preserving all other positions. We consider randomized algorithms that release a real answer after every update.

\begin{definition}
Two probability distributions $P,P'$ are {\em $(a,b)$-close}, denoted as $P\approx_{a,b}P'$, if every measurable event $E$
satisfies $P(E)\le e^a P'(E)+b$ and $P'(E)\le e^aP(E)+b$.
We use the same notation for random variables, comparing their distributions.
\end{definition}

\begin{definition}[Differential privacy under continual observation~\cite{DNPR10,CSS11}]
A randomized algorithm $A$ is \emph{continually $(\eps,\delta)$-DP} if for every pair of neighboring streams $\mathbf u,\mathbf u'$ we have
\[
 A_{1:T}(\mathbf u)\approx_{\eps,\delta}
 A_{1:T}(\mathbf u').
\]
\end{definition}

\begin{definition}[Retroactivity~{\cite{DIL07}}]
We say that two streams $\mathbf u,\mathbf u'$ \emph{coalesce} at time $b$ if
$D_t(\mathbf u)=D_t(\mathbf u')$ for all $t\ge b$.
A randomized algorithm $A$ has \emph{$(\rho,\eta)$-retroactivity} if for every $b\in[T]$ and every pair of streams coalescing at $b$ we have
\begin{equation*}
 A_{b:T}(\mathbf u)\approx_{\rho,\eta}
 A_{b:T}(\mathbf u').
\end{equation*}
This compares the whole suffix
without conditioning on earlier answers. Exact retroactivity is the
case where $\rho=\eta=0$.
\end{definition}

\begin{definition}
An algorithm $A$ is $(\Delta,\beta)$-accurate w.r.t.\ a scalar function $F$ with time horizon $T$ if for every fixed input stream $\mathbf u$, with probability at least $(1-\beta)$, it guarantees simultaneous error at most $\Delta$. That is,
\[
 \Prb\!\left[\max_{t\le T}|A_t-F(D_t)|\le\Delta\right]\ge1-\beta.
\]
\end{definition}

\subsection{Our Results}
\label{sec:problem}

We consider the following problem. For a public horizon $T\ge8$, let
$d=\lfloor T^{1/3}\rfloor$ be the size of the point universe.
Each record in the stream is either a point $x\in[d]$
or a query $Q\subseteq[d]$.
A query is represented by a $d$-bit membership vector. At time $t$,
let $S_t$ be the multiset of points present in $D_t$, and let
$c_Q(t)$ be the multiplicity of the query $Q$ in $D_t$.
The goal is to maintain a running approximation to the following function $F$, which we call the \emph{\textsf{CappedCount}} function.
\begin{equation*}
 \boxed{F(D_t)=\max_{Q\subseteq[d]}
       \frac{\min\{c_Q(t),\,|S_t\cap Q|\}}{\sqrt{1+|S_t|}}.}
\end{equation*}
Here intersection sizes are counted with multiplicities. 

\medskip
We show the following results.

\begin{theorem}[Positive result, informal]\label{intro:positive}
There is a continually-DP algorithm for \textsf{CappedCount} 
 with simultaneous error
$O({\rm polylog}(T))$. 
\end{theorem}

\begin{theorem}[Negative result, informal]
Every continually-DP and retroactive algorithm for \textsf{CappedCount} requires worst-case error
$\Omega({\rm poly}(T))$, even if accuracy is only required to hold separately
at each time with constant probability.
\end{theorem}

\subsection{Technical overview}

\paragraph{Positive result.}
Throughout the execution, we process {\em point} insertions and deletions using a  dynamic variant of the Private Multiplicative Weights (PMW) algorithm~\cite{HR10,QY25}, supporting evolving datasets. We also use a private counter to estimate the current number of points $|S_t|$. We process {\em query} insertions and deletions using a continually-DP sparse histogram to track the multiplicity $c_Q(t)$ of all ``noticeable'' queries $Q$ whose current multiplicities are large enough to matter. In every time step, we feed every such noticeable query $Q$ to the PMW to obtain an estimate of its answer $|S_t\cap Q|$. The error in this estimation scales as $\sqrt{|S_t|}$, and so the normalization by (our estimation of) $\sqrt{|S_t|}$ leaves polylogarithmic error. Intuitively, privacy for the points comes from the dynamic PMW
and the private counter, while privacy for the query records comes from the sparse histogram.

\paragraph{Negative result.}
We build on an impossibility result of Dwork, Naor, and Vadhan~\cite{DNV12}, who showed that {\em stateless} DP algorithms for answering many queries must incur large error. To see the connection, suppose we had a continually-DP and retroactive algorithm $A$ for \textsf{CappedCount} with small error. We could transform it into a ``weakly stateful'' DP algorithm for answering queries arriving online as follows:
\begin{enumerate}
    \item Take an input dataset $X$ of $N$ points from a universe $[d]$, and insert its points one-by-one into $A$'s stream.
    \item When given a query $Q$ (represented as a subset $Q\subseteq[d]$): 
    \begin{itemize}
        \item Insert $N$ copies of the query record $Q$ into $A$'s stream, and let $a$ be $A$'s answer after the last insertion.
        \item Delete all these $N$ query records from the stream.
        \item Return the (unnormalized) answer $a\,{\cdot}\,\sqrt{N+1}$ as the estimate of $|X\cap Q|$.
    \end{itemize}
\end{enumerate}
By the retroactivity requirement, the answer to the $j$-th query should be almost independent of all past queries, which intuitively (but somewhat inaccurately) seems very close to the behavior of a stateless algorithm. Hopefully, we could then apply \cite{DNV12}'s negative result to rule out such an algorithm $A$.

The issue, of course, is that the resulting algorithm for answering queries is {\em not} stateless, as retroactivity allows stateful algorithms. To emphasize the difference:
\begin{itemize}
    \item With a stateless algorithm, as in \cite{DNV12}, once the dataset, the current query, and the algorithm's coins are fixed, the $j$-th answer does not depend on previous queries.
    \item In contrast, with a retroactive algorithm, if we fix its coins, then the answer to the $j$-th query may depend on past deleted queries (via the algorithm's internal state). This happens because the retroactivity requirement need only hold when {\em averaging over the algorithm's coins}. 
\end{itemize}

To obtain our negative result, we therefore prove a lower bound for such ``weakly stateful'' DP algorithms, building on the techniques of~\cite{DNV12}. We obtain the following result:

\begin{definition}[Prefix forgetting]\label{intro:def:prefix}
A randomized (potentially stateful) mechanism $M$ that takes a dataset and answers a sequence of queries is {\em $(\rho,\eta)$-prefix forgetting} if the following holds for every dataset $X$, every sequence of queries $\mathbf Q=(Q_1,\ldots,Q_k)$, and every $b\in\{0,\ldots,k\}$.  Let
$\mathbf Q^{(b)}$ be the same as $\mathbf Q$ after replacing its first $b$ queries by the dummy query $\bot$. Then,
\begin{equation*}
 M_{b+1:k}(X,\mathbf Q)
 \approx_{\rho,\eta}
 M_{b+1:k}(X,\mathbf Q^{(b)}).
\end{equation*}
\end{definition}

\begin{theorem}[Prefix-forgetting lower bound, informal]\label{intro:DNV-extended}
For all sufficiently large $N$, there
are a universe of size $d=\Theta(N^{3/2})$ and a query horizon
$k=\Theta(N^{3/2}\log N)$ for which every $(1,1/N)$-DP mechanism with $(1,1/N)$-prefix forgetting
requires worst-case unnormalized error $\Omega(N^{3/4})$
when answering subset queries on an $N$-record database.
\end{theorem}

Our reduction outlined above converts a private and retroactive algorithm with error $\Delta$ into a prefix-forgetting mechanism for answering queries with (unnormalized) error
$\Delta\sqrt{N+1}$. In our setting the universe size is $d\approx T^{1/3}\approx N^{3/2}$, so  $N\approx T^{2/9}$. Thus, applying Theorem~\ref{intro:DNV-extended} gives $\Delta=\Omega(N^{1/4})=\Omega(T^{1/18})$.

\medskip
Prefix forgetting is related to the \emph{many-to-one analyst-privacy} model of~\cite{DNV12}. In this model, several different analysts interact with the same query answering mechanism that holds a dataset $X$. Each analyst only sees the answers to its own
queries. There are two privacy requirements: 
\begin{enumerate}
    \item[{\bf (1)}] {\bf Privacy for the data.} We need to preserve DP w.r.t.\ the dataset $X$ as usual;
    \item[{\bf (2)}] {\bf Privacy for the analysts.} The queries issued by the analysts should not leak to other analysts. Formally, the requirement, which \cite{DNV12} calls {\em many-to-one}, is that one analyst's entire view remain
$(\rho,\eta)$-close when all other analysts change their query
strategies, while that analyst's strategy is held fixed.
\end{enumerate}
Note that this is similar to prefix forgetting as both notions require insensitivity to modifying a large subset of the queries in the stream: in prefix forgetting the subset is a prefix of queries, and in many-to-one the subset is the queries given by all other analysts (which may be interleaved with those of the protected analyst).

Dwork et al.~\cite{DNV12} gave
an algorithm in their many-to-one
analyst-privacy model with normalized error $\widetilde O(N^{-1/4})$. In contrast, without analyst-privacy, error $\widetilde O(N^{-1/2})$
is achievable~\cite{HR10}. Dwork et al.~\cite{DNV12} asked whether this gap in accuracy
is inherent. The following corollary settles the dependence on $N$ up to logarithmic factors. 

\begin{corollary}[Many-to-one analyst-privacy]
\label{cor:analyst-lower}
Fix constants $\eps,\rho\ge0$. There exists a constant $C$ such that
the following holds for all sufficiently large $N$, every universe
of size $|U|\ge CN^{3/2}$, and every query horizon
$k\ge CN^{3/2}\log(N+1)$. Let $M$ be an online mechanism that
supports at least $CN$ analyst IDs, whose entire answer transcript
is $(\eps,1/N)$-DP under replacement of one data record, and which
provides $(\rho,1/N)$ many-to-one analyst-privacy. Suppose that, for every fixed dataset
$X\in U^N$ and list $(\mathrm{id}_j,Q_j)_{j=1}^k$ of subset queries
with analyst IDs, its answer at each position $j$ estimates
$|X\cap Q_j|/N$ with error at most $\alpha$ and success probability
at least $0.99$.
Then $\alpha=\Omega(N^{-1/4})$.
\end{corollary}

We prove Corollary~\ref{cor:analyst-lower} in
Appendix~\ref{app:querylower}. The idea behind this corollary is
that the proof of Theorem~\ref{intro:DNV-extended} organizes the
queries into $O(N)$ consecutive blocks and uses prefix forgetting
only at block boundaries, and only for the joint answers within
the next block. Thus, to obtain the corollary, we assign each
block to a different analyst. Many-to-one analyst-privacy then
provides exactly the comparison needed: the current analyst's
view must be approximately insensitive to all previous analysts
replacing their queries by empty-set queries.

\subsection{Additional Related Works}
Several other works studied the analyst-privacy model. In particular, Hsu, Roth, and Ullman~\cite{HRU13} presented constructions achieving error
$\widetilde O(N^{-1/2})$ for one-query-to-many analyst-privacy and
$\widetilde O(N^{-1/3})$ for one-analyst-to-many analyst-privacy.
These notions protect one query, or one analyst's workload, against
the joint view of all other analysts. Many-to-one analyst-privacy (the notion which our work has implications to) instead
protects all other analysts' workloads simultaneously against one
analyst. Later work studies budget sharing and fairness across
analysts~\cite{PWFM21,PSFM22}, and privacy accounting for each
analyst~\cite{ZH23}; these works concern data privacy rather than
hiding query workloads.

\section{A Positive Result for Continual-DP}
\label{sec:positive}
\label{sec:dponly}
\label{sec:upper}

In this section we formalize and prove Theorem~\ref{intro:positive}. Specifically, we show the following theorem.

\begin{theorem}[Continual DP without retroactivity]
\label{thm:dp-upper}
For all sufficiently large $T$, there is a continually
$(1,T^{-100})$-DP algorithm for \textsf{CappedCount} over the horizon $T$,
with simultaneous error $O(\log^5T)$ and failure probability at most
$T^{-100}$.
\end{theorem}

\begin{observation}
The add/remove sensitivity of the \textsf{CappedCount} function is at most one.
\end{observation}

\begin{proof}
Consider inserting one record into $D_t$.
The sensitivity w.r.t.\ inserting a query record $Q$ is trivially
bounded by 1, as it changes the numerator (for the query $Q$) by at
most one and leaves the denominator unchanged.

As for inserting a point $x$, consider the score of an arbitrary
query $Q$ and let $z\le|S_t|$ denote its numerator. The insertion
increases the point count from $|S_t|$ to $|S_t|+1$ (thereby increasing the denominator), and can
either increase the numerator by 1 or leave it unchanged. If the
numerator remains unchanged, then the score decreases by
\[
 \frac{z}{\sqrt{1+|S_t|}}-\frac{z}{\sqrt{2+|S_t|}}
 =\frac{z\bigl(\sqrt{2+|S_t|}-\sqrt{1+|S_t|}\bigr)}
       {\sqrt{(1+|S_t|)(2+|S_t|)}}
 \le\frac{|S_t|}{1+|S_t|}\le1 .
\]
If the numerator increases, then the score increases by
\[
 \frac{z+1}{\sqrt{2+|S_t|}}-\frac{z}{\sqrt{1+|S_t|}}
 \le\frac{z+1}{\sqrt{2+|S_t|}}-\frac{z}{\sqrt{2+|S_t|}}
 =\frac{1}{\sqrt{2+|S_t|}}\le1 .
\]
So the change is bounded by 1 either way.
\end{proof}

As we mentioned in the introduction, in our continually-DP algorithm for \textsf{CappedCount} we will privatize points (additions and deletions) in the stream using a dynamic variant of the PMW, and privatize the query records throughout the stream using a dynamic continually-DP sparse histogram. For the dynamic variant of the PMW we will use the following construction by Qiu and Yi \cite{QY25}. 
At a high level, their construction combines the ordinary PMW algorithm~\cite{HR10} with the binary-tree technique for continual observation, where queries are answered by adding and subtracting estimates computed by PMW instances on snapshots and blocks of deleted points. 
We include the details in
Appendix~\ref{app:qy} for completeness.

\begin{lemma}[restatement of {\cite[Theorem~8]{QY25}}]
\label{lem:onlineqy}
Let $\mathcal U$ be a finite value universe, let $T$ be the time
horizon, and let $K$ be a bound, known in advance, on the number of
queries. An initially empty multiset of values from $\mathcal U$
undergoes unit insertions, deletions, and no-ops over $T$ steps, where each
deletion names the insertion time of the occurrence it removes, as in
Section~\ref{sec:model}. Let $S_t$ denote this multiset after the
$t$-th update. In addition, throughout the execution at most $K$ subset-queries $Q\subseteq\mathcal U$ arrive online. The goal is to answer every such query $Q$ (at the time $t$ at which it arrives) with an answer $\widehat a_Q(t)$ that should approximate its true answer $|S_t\cap Q|$. 

For every $0<\beta,\eps,\delta\le1$ there is a continually $(\eps,\delta)$-DP mechanism such that,
with probability at least $1-\beta$, all of its answers throughout the execution simultaneously
satisfy
\begin{equation}
 \Bigl|\widehat a_Q(t)-|S_t\cap Q|\Bigr|
 = O\!\left(\left(\sqrt{\frac{|S_t|}{\eps}}
      +\frac1\eps\log^3\frac{|\mathcal U|+T+K}{\delta\beta}\right)
   \log^5\frac{|\mathcal U|+T+K}{\delta\beta}\right).
 \label{eq:onlineqy}
\end{equation}
The mechanism is {\em oblivious} in the sense that it guarantees utility for every fixed sequence of point additions/deletions and queries. The privacy guarantee is w.r.t.\ every pair of such sequences with the same public queries and arrival times, whose update streams differ by replacing one point update with a no-op, as in Section~\ref{sec:model}.
\end{lemma}

We will use standard techniques for continually-DP sparse histograms. This can be achieved, informally, using the binary-tree technique by maintaining noisy histograms at its nodes and by combining them to estimate the current counts.

\begin{lemma}[Dynamic sparse histograms,
  {e.g.,~\cite{RCR22,KKMNS26}}]
\label{lem:sparsehist}
Let $\mathcal U$ be a finite value universe and let $T$ be a public
time horizon. An initially empty multiset of values from $\mathcal U$
undergoes unit insertions, deletions, and no-ops over $T$ steps,
where each deletion names the insertion time of the occurrence it
removes, as in Section~\ref{sec:model}. Let $S_t$ denote this multiset
after the $t$-th update, and let $c_x(t)$ denote the multiplicity
of $x\in\mathcal U$ in $S_t$.

For every $0<\eps\le1$ and $0<\delta,\beta<1$, there is a 
continually $(\eps,\delta)$-DP mechanism that, at each time $t$,
releases a set $R_t\subseteq\mathcal U$ and an estimate
$\widehat c_x(t)$ for each $x\in R_t$. Define
$\widehat c_x(t)=0$ for $x\notin R_t$. The mechanism satisfies:
\begin{enumerate}
    \item On every execution and at every time,
    $R_t\subseteq\{x\in\mathcal U:c_x(t)>0\}$. In particular, $|R_t|\le |S_t|\le T$.

    \item For every fixed update stream, with probability at least
    $1-\beta$, simultaneously for every $t\in[T]$ and
    $x\in\mathcal U$, it holds that  $|\widehat c_x(t)-c_x(t)|\le O\!\left(
          \frac{\log^2(2T)}{\eps}
          \log\frac{2T}{\delta\beta}
       \right)$.
    
\end{enumerate}
The privacy guarantee is w.r.t.\ the replacement of one update by a no-op.
\end{lemma}

\begin{remark}
    To simplify the presentation, we have stated Lemmas~\ref{lem:onlineqy} and~\ref{lem:sparsehist} for the {\em oblivious} setting. Standard arguments show that, in fact, both constructions carry over to the adaptive setting where future inputs may be chosen based on previous outputs of the mechanism.
\end{remark}

We are now ready to prove Theorem~\ref{thm:dp-upper}.

\begin{proof}[Proof of Theorem~\ref{thm:dp-upper}]
Denote 
$m(t)=|S_t|$, $a_Q(t)=|S_t\cap Q|$, and $\delta =\beta =T^{-100}/10$, where $T$ is the public horizon. The three primitives use independent randomness. Our continually-DP algorithm for \textsf{CappedCount} works as follows.

\begin{enumerate}
    \item Initialize the following primitives:
    \begin{itemize}[itemsep=0px, topsep=0px]
        \item $\mathsf{Hist}$: A continually-DP sparse histogram, as in Lemma~\ref{lem:sparsehist}, with privacy parameters $(\frac{1}{3}{,\delta})$ and failure probability $\beta$.
        \item $\mathsf{Count}$: Binary-tree counter satisfying $(1/3,0)$-continual DP, as in  \cite{DNPR10}.
        \item $\mathsf{PMW}$: A fully dynamic mechanism for answering queries, as in Lemma~\ref{lem:onlineqy}, with privacy parameters $(\frac{1}{3}{,\delta})$ and failure probability $\beta$ and with $\mathcal U=[d]$ and $K=T^2$. 
        \textcolor{gray}{{\small \% There will be at most $T$ queries per time step, and hence at most $K=T^2$ queries throughout the execution.}}
    \end{itemize}
    \item For $t=1,2,\dots,T$ do:
    \begin{enumerate}[itemsep=0px, topsep=0px]
        \item Read the update $u_t$.
\item If $u_t$ contains a {\em query} update, then send it to $\mathsf{Hist}$; otherwise send a no-op to $\mathsf{Hist}$.
\item If $u_t$ contains a {\em point} update, then send it to $\mathsf{PMW}$; otherwise send a no-op to $\mathsf{PMW}$.
\item Feed $\mathsf{Count}$ the increment $+1$ for a point insertion, $-1$ for an effective point deletion, and $0$ otherwise.
\item Obtain from $\mathsf{Hist}$ a set $R_t$ of all ``noticeable'' queries together with non-negative estimates for their counts $\{\widehat c_Q(t):Q\in R_t\}$. \textcolor{gray}{{\small \% We can assume that $\widehat c_Q(t)$ are non-negative by clipping negative values to 0.}}
\item Let $\widehat m(t)$ denote $\mathsf{Count}$'s current  estimate, after clipping it to $[0,T]$.
\item Query $\mathsf{PMW}$ on each $Q\in R_t$, in a fixed public order, to obtain an approximate answer $\widehat a_Q(t)$, clipped to $[0,\widehat m(t)]$. \textcolor{gray}{{\small \% The set $R_t$ is computed by $\mathsf{Hist}$ independently of $\mathsf{PMW}$'s answers.}}
\item\label{step:computeMax} If $R_t=\emptyset$ or $\widehat m(t)\leq H=\Theta\left(\log^6 T\right)$ then set $y_t\gets0$. Otherwise set
$$
y_t\leftarrow\max_{Q\in R_t}\left\{
 \frac{\min\{\widehat c_Q(t),\,
       \widehat a_Q(t)\}}
      {\sqrt{1+\widehat m(t)}}
\right\}
$$
\item Release $y_t$.
    \end{enumerate}
\end{enumerate}

\paragraph{Privacy.}
The algorithm's output transcript is postprocessing of the joint
transcripts of $\mathsf{Hist}$, $\mathsf{PMW}$, and $\mathsf{Count}$,
so adaptive composition gives $(1,2\delta)$-continual DP and hence
the claimed $(1,T^{-100})$ guarantee.

\paragraph{Utility.}
Let $h=\Theta(\log^3T)$ and $H=\Theta(\log^6T)$ be sufficiently large with $H\ge2h$. 
By the accuracy guarantees of $\mathsf{Hist}$, $\mathsf{Count}$, and $\mathsf{PMW}$, with probability at least $1-3\beta\geq1-T^{-100}$, simultaneously at all times $\mathsf{Hist}$'s and $\mathsf{Count}$'s answers are accurate to within error $h$, and $\mathsf{PMW}$'s answers are accurate to within error
$E(t)=O\left(\sqrt{m(t)+H}\,\log^5T\right)$.

Assume first that $\widehat m(t)> H$. Then, as $\left|m(t)- \widehat m(t)\right|\leq h$, we have that $m(t)$ and
$\widehat m(t)$ are within constant factors. In this case, for every query $Q$, our estimation error is of the order of
\begin{equation*}
O\!\left(\frac{h+E(t)}{\sqrt{1+m(t)}}\right)
=
 O(\log^5T),
\end{equation*}
and thus the same error bound holds under the maximum over $Q$ computed in Step~\ref{step:computeMax}.

Otherwise, if $\widehat m(t)\leq H$, then the true (noiseless) quantity for every $Q$ is
\begin{equation*}
\frac{\min\{c_Q(t), a_Q(t)\}}{\sqrt{1+m(t)}}
\leq \frac{m(t)}{\sqrt{1+m(t)}}
\leq \sqrt{m(t)}
\leq \sqrt{H+h}
=O(\log^3T),
\end{equation*}
and outputting zero is accurate enough.
\end{proof}

\section{A Negative Result for Continual-DP and Retroactivity}
\label{sec:negative}

In this section we formalize and prove our negative result. Specifically, we show the following theorem.

\begin{theorem}[Continual DP with retroactivity]
\label{thm:retro-lower}
For all sufficiently large $T$, 
every continually $(1,1/T)$-DP and $(1,1/T)$-retroactive algorithm for the \textsf{CappedCount} function requires worst-case error
$\Omega(T^{1/18})$, even if accuracy is required only separately
at each time with success probability $0.99$.
\end{theorem}

As we mentioned in the introduction, our proof builds on (an extension of) a negative result
by Dwork, Naor, and Vadhan~\cite{DNV12}. Let us start by recalling the setting for this negative result.
For a dataset $X=(x_1,\ldots,x_N)\in U^N$, the normalized answer to
a subset query $Q\subseteq U$ is $|X\cap Q|/N$.
Queries arrive online, and each is answered before the next arrives.
We say that a mechanism $M$ is $(\alpha,0.01)$-accurate if, for every
fixed dataset and list of subset queries, its answer at each position
$j$ satisfies $\Prb\!\left[\left|M_j-\frac{|X\cap Q_j|}{N}\right|\le\alpha\right]\ge0.99$.
Privacy is w.r.t.\ replacing one row of $X$ and is required for the
entire answer transcript. Dwork et al.~\cite{DNV12} showed a negative result for {\em stateless} mechanisms in this context. We extend their negative result to hold also for stateful algorithms, provided they are {\em prefix forgetting}, as in Definition~\ref{intro:def:prefix}.

\begin{theorem}[Extension of \cite{DNV12}]
\label{thm:querylower}
Fix constants $\eps,\rho\ge0$. There exists a constant $C$ such that the following holds for every sufficiently large $N$, every universe of size $|U|\ge C N^{3/2}$, and every query horizon $k\ge C N^{3/2}\log(N+1)$. Let $M$ be an $(\eps,\frac{1}{N})$-DP and $(\rho,\frac{1}{N})$-prefix forgetting mechanism with accuracy $(\alpha,0.01)$. Then $\alpha=\Omega(N^{-1/4})$.
\end{theorem}

We prove Theorem~\ref{thm:querylower} in Appendix~\ref{app:querylower}.
We now use it to prove Theorem~\ref{thm:retro-lower}.

\begin{proof}[Proof of Theorem~\ref{thm:retro-lower}]
Suppose that $A$ is continually $(1,1/T)$-DP and
$(1,1/T)$-retroactive, and has
error at most $\Delta$ with probability at least $0.99$ at every time
of every fixed stream.
Let $C$ be the constant from Theorem~\ref{thm:querylower} for
$(\eps,\rho)=(2,1)$.
Choose $N=\Theta(T^{2/9})$, with a sufficiently small implicit
constant so that $d=\lfloor T^{1/3}\rfloor=\Theta\left(N^{3/2}\right)$ is at least the
required size of the hard universe $U$ in
Theorem~\ref{thm:querylower}.
We may therefore take $U=[d]$, the point universe in the
definition of \textsf{CappedCount}.
Let $k=\lceil C N^{3/2}\log(N+1)\rceil$.

We use $A$ to construct the following prefix forgetting mechanism $M$ for answering $k$ queries given a dataset $X\in U^N$:
\begin{enumerate}
    \item Initialize $A$ and insert the $N$ point records
    $x_1,\ldots,x_N$ one by one.
    \item Upon receiving a query $Q_j$:
    \begin{enumerate}[itemsep=0px, topsep=0px]
        \item If $Q_j=\bot$, feed $A$ $2N$ no-ops and return zero.
        \item Otherwise, insert $N$ copies of the query record $Q_j$ one by one,
        and let $y_j$ be $A$'s output after the last insertion.
        \item Delete these $N$ query records one by one, specifying
        their insertion times.
        \item Return
        \[
        \widehat a_j=\clip_{[0,1]}\!\left(
             \frac{\sqrt{N+1}}N y_j\right).
        \]
    \end{enumerate}
\end{enumerate}
Every query uses exactly $2N$ stream positions, and the simulation
fits into the public horizon because
\[
 N+2Nk=O(T^{5/9}\log T)\le T
\]
for sufficiently large $T$; unused positions are filled with no-ops.

\paragraph{Privacy.}
The mechanism $M$ simply post-processes $A$'s answers, so DP
w.r.t.\ $X$ follows from closure under post-processing.
More specifically, a replacement of one row of $X$ corresponds
to two neighboring-steps for $A$, via a no-op.\footnote{For
$A$, neighboring streams differ by replacing one operation with
a no-op; for $M$, neighboring datasets differ by replacing one row.}
Thus, group privacy gives $(2,(1+e)/T)$-DP for $M$, and hence
$(2,1/N)$-DP for sufficiently large $T$.

\paragraph{Prefix forgetting.}
Replacing a query prefix by dummies replaces all of its query updates
by no-ops while preserving every subsequent update and timestamp.
Before the next query, both streams contain exactly the same point
records with the same insertion times and no query records, so they
coalesce, and retroactivity of $A$ gives $(1,1/T)$-prefix
forgetting for $M$ by postprocessing, with $1/T\le1/N$.

\paragraph{Utility.}
When $y_j$ is recorded, the point dataset is $X$, and the only active
query type is $Q_j$, with multiplicity $N$, so the true scalar value is
\[
 \frac{\min\{N,|X\cap Q_j|\}}{\sqrt{N+1}}
 =\frac{|X\cap Q_j|}{\sqrt{N+1}}.
\]
Thus, for every fixed dataset and list of queries, the
answer $\widehat a_j$ has error at most
\[
 \alpha=\frac{\Delta\sqrt{N+1}}N
\]
with probability at least $0.99$ at each query position.
Theorem~\ref{thm:querylower} gives $\alpha=\Omega(N^{-1/4})$, and hence
\[
 \Delta=\frac{N}{\sqrt{N+1}}\,\alpha
        =\Omega(N^{1/4})
        =\Omega(T^{1/18}),
\]
which proves the claim.
\end{proof}

\paragraph*{AI Disclosure.}
We used ChatGPT (OpenAI), Claude (Anthropic), and Gemini (Google) to assist in preparing this paper. For producing initial drafts of some arguments, for drafting and revising the exposition, and for formalizing and auditing proofs. The tools were used interactively and materially affected the results and writing throughout. The authors verified the correctness and originality of all content, and take full responsibility for all claims.

\clearpage
\appendix
\section{Details of the online-query formulation of Qiu--Yi}
\label{app:qy}

In this appendix we include a proof of Lemma~\ref{lem:onlineqy}. The
construction and the analysis are due to Qiu and Yi~\cite{QY25}. We
include the details here only for completeness, as we needed to restate their theorem to handle queries that arrive online (they stated it for a query class fixed in advance).

We first assume that every
deletion removes a present occurrence $(x,\tau)$, specified by its
value and insertion time. We call such streams \emph{valid}, and
consider neighboring valid streams that differ by replacing one
operation with a no-op. We handle ineffective deletions at the end.
Denote
\[
 \varepsilon_0=\eps/2,\qquad
 \delta_0=\frac{\delta}{1+e^{\eps/2}},\qquad
 \ell=\lceil\log_2(2T)\rceil,\qquad
 B_{\max}=2^{\lceil\log_2T\rceil}.
\]
All binary trees below use the public clock $[T]$, including no-ops.

We start with the decomposition of~\cite[Definition~5 and Lemma~6]{QY25}.
For every $v\in[T]$, let $h(v)$ be the largest power of two dividing
$v$, and let $I_v=\{v-h(v)+1,\ldots,v\}$. At time $v$, create a collection
containing the currently present occurrences inserted in $I_v$.
This collection subsequently only loses points. Let $S_v(t)$ be
its contents at time $t\ge v$, and write $m_v(t)=|S_v(t)|$.
Starting with $v=t$ and repeatedly replacing $v$ by $v-h(v)$ gives
a cover $\mathcal C(t)$ with at most $\ell$ nodes, satisfying
\begin{equation}
 S_t=\mathop{\biguplus}_{v\in\mathcal C(t)}S_v(t).
 \label{eq:qy-prefix-partition}
\end{equation}
Here $\biguplus$ denotes union that keeps multiplicities.
Indeed, every point present at time $t$ was present when its covering
collection was created. At each tree level the insertion intervals
are disjoint, so a changed occurrence affects at most one collection
per level.

We maintain query answers for these collections as follows. Choose
a sufficiently large universal constant $C$ and set
\begin{equation*}
 \beta_* =\frac{\beta}{(T+1)^C},\qquad
 \Gamma=\frac{C\ell^3}{\varepsilon_0}\log\frac{2T}{\beta_*}.
\end{equation*}
\begin{enumerate}
    \item For every collection $S_v$, initialize a binary-tree
    counter for its size. Its input is $m_v(v)$ at time $v$,
    followed by $-1$ for every deletion from the collection and
    zeros elsewhere. Add independent $\Lap(C\ell^2/\varepsilon_0)$
    noise to every tree node, and let $\widehat m_v(t)$ denote
    the resulting size estimate.

    \item At every time $t\ge v$, compute
    \begin{equation*}
     b_v(t)=2^{\left\lceil\log_2
       \clip_{[1,B_{\max}]}(\widehat m_v(t)+\Gamma)\right\rceil},
     \qquad B_v(t)=\min_{v\le s\le t}b_v(s).
    \end{equation*}
    Start a new epoch at time $v$ and whenever $B_v(t)$ decreases.
    There are at most $\ell$ epochs, since $B_v(t)$ takes values
    in $\{1,2,4,\ldots,B_{\max}\}$ and never increases.
    At the start $s$ of an epoch, store the snapshot $S_v(s)$
    and initialize a binary tree for subsequent deletions from
    this snapshot. The tree uses fixed global-time intervals,
    with zeros at times at most $s$ and whenever no snapshot
    point is deleted. Blocks ending by $s$ are empty and answered
    exactly; blocks crossing $s$ include these leading zeros.

    \item Maintain a separate persistent online PMW
    instance~\cite{HR10} for every snapshot and completed deletion
    block, with replacement-privacy parameters
    \begin{equation*}
     (\varepsilon_s,\delta_s)=
     \left(\frac{\varepsilon_0}{C\ell^3},
           \frac{\delta_0}{C\ell^3}\right)
    \end{equation*}
    and failure probability $\beta_*$. If the epoch's capacity
    is $B$, keep the first $B$ occurrences of the static database
    in a fixed order and pad it with dummy records to size $B$.
    Use the universe
    $\mathcal V=(\mathcal U\times[T])\cup\{\star\}$, lifting a
    subset query $Q\subseteq\mathcal U$ to $Q\times[T]$, which
    excludes the dummy $\star$.
    Initialize each needed PMW instance only after its static
    database is complete, and feed it queries in their arrival order.

    \item To answer a query at time $t$, consider every
    $v\in\mathcal C(t)$. Query the current snapshot and the
    $O(\ell)$ nodes covering its deletion prefix. Subtract the
    deletion answers from the snapshot answer, and sum the
    resulting estimates over $v\in\mathcal C(t)$.
\end{enumerate}
Each PMW instance receives at most $K$ queries. The online PMW accuracy
analysis applies to these submitted queries, so their descriptions need
not be known in advance. A constant rescaling of the privacy
budget converts PMW's add/remove guarantee to replacement privacy.

\paragraph{Privacy.}
Fix the external query schedule. A neighboring change alters at most
two entries of a collection's count stream by one, so the noise above
gives its counter privacy parameter $\varepsilon_0/(2\ell)$.
Parallel composition within each level, followed by composition
across levels, gives total counter privacy $\varepsilon_0/2$.

Fix an arbitrary sequence $z$ of counter outputs. This determines the
epoch boundaries, capacities, and queries sent to each PMW instance.
A changed occurrence affects at most one collection for each interval
length, and at most $\ell$ epochs per collection. Within each epoch, it
affects one snapshot and at most one block per level of the deletion tree. Thus it affects
$O(\ell^3)$ static databases. After truncating
and padding, each changes by at most one replacement, for every $z$.
Composition therefore gives $(\varepsilon_0/2,\delta_0)$-DP for
their joint transcript. The counters and PMW instances use independent
randomness, so composition gives $(\varepsilon_0,\delta_0)$-DP for
the mechanism, even though their execution is interleaved.
The same argument permits composition with a DP mechanism that chooses
the query schedule using its own independent randomness and without
access to this mechanism's answers or random coins.

\paragraph{Utility.}
The standard binary-tree counter bound gives error at most $\Gamma$
throughout each counter's lifetime except with probability $\beta_*$.
There are at most $T$ counters and $O(T^2\ell)$ PMW instances.
When applying PMW accuracy, fix the counter transcript and then
average over it. A union bound therefore shows that all accuracy
guarantees hold simultaneously with probability at least $1-\beta$.

Assume that these guarantees hold. Since each $m_v(t)$ only
decreases, at every time we have
\begin{equation*}
 m_v(t)\le B_v(t)\le2\max\{1,m_v(t)+2\Gamma\}.
\end{equation*}
For the lower bound, every earlier $b_v(s)$ is at least
$m_v(s)\ge m_v(t)$. For the upper bound, use $B_v(t)\le b_v(t)$.
Thus no snapshot is truncated when its epoch starts, and the
deletion blocks, which are subsets of that snapshot, are not
truncated either.

Denote $\Lambda=\log((|\mathcal U|+T+K)/(\delta\beta))$,
$H=C_0\Lambda^6/\eps$, and $m(t)=|S_t|$, for a sufficiently large
constant $C_0$. Then $\Gamma=O(\Lambda^4/\eps)\le H$ and
$\varepsilon_s=\Omega(\eps\Lambda^{-3})$.
Every PMW instance queried at time $t$ has capacity
$B=O(m(t)+H)$. Also, $B\le2T$ and
$|\mathcal V|=|\mathcal U|T+1$, so the logarithms of
$|\mathcal V|+1$, $2/\delta_s$, $K+1$, and $B/\beta_*$ are
$O(\Lambda)$. Multiplying the normalized PMW error
in~\cite[Theorem~4.15]{DR14} by $B$ gives count error
\begin{equation*}
 O\!\left(\sqrt{\frac B{\varepsilon_s}}\,\Lambda\right)
 =O\!\left(\sqrt{\frac{m(t)+H}{\eps}}\,\Lambda^{5/2}\right).
\end{equation*}
If the public PMW bound computed from $B$ exceeds $B$, the instance returns zero
instead, which has error at most $B$.
Each answer combines $O(\ell^2)$ PMW estimates, so summing their
errors gives~\eqref{eq:onlineqy}, using~\eqref{eq:qy-prefix-partition}.

\paragraph{Effective deletions.}
\label{app:effective-deletions}
We now remove the assumption that the stream is valid. Replace
every ineffective deletion request by a no-op and apply the
construction above. This preserves the current database at every time.
Neighboring original streams become valid streams at distance at
most two. Omitting an insertion removes it and its first effective
deletion; a valid path removes the deletion first. Omitting an
effective deletion removes it or moves it to the next deletion
request for that occurrence; a valid path removes the earlier
deletion before adding the later one. Omitting an ineffective request
changes nothing. Group privacy therefore gives
$(2\varepsilon_0,(1+e^{\varepsilon_0})\delta_0)=(\eps,\delta)$-DP
for the original streams, completing the proof.

\section{Proof of the prefix-forgetting extension of DNV}
\label{app:querylower}

In this appendix we prove Theorem~\ref{thm:querylower}. We build on
Dwork, Naor, and Vadhan's attack on stateless
mechanisms~\cite[Theorem~II.7 and Proposition~II.5]{DNV12}. Their
queries are fixed on all but one database row and uniformly random
elsewhere. We instead hide a constant fraction of the rows in each
of several groups. Accuracy and prefix forgetting force the answers
to be correlated with the hidden records, while differential privacy
gives a contradictory upper bound on this correlation.
The database and the entire query list are sampled before the
mechanism's coins. We use accuracy only in the original execution;
comparisons with modified executions use privacy and prefix forgetting.

\paragraph{Roadmap.} Section~\ref{sec:ordinary-encoding} turns the answers to each block of subset queries into a decoded vector $Y$ that approximates the vector $\mu$ of group means. Section~\ref{sec:ordinary-source} defines the hard input, in which $\mu$ depends on fresh signs given to hidden rows. Section~\ref{sec:ordinary-score} combines the two: since $Y$ is close to $\mu$ and $\mu$ is correlated with the hidden signs, accuracy and prefix forgetting force $Y$ to be correlated with them as well. Section~\ref{sec:ordinary-soundness} shows that differential privacy forces this correlation to be small, and Section~\ref{sec:ordinary-parameters} chooses parameters under which the two bounds contradict each other.

\subsection{Robust encoding by subset queries}
\label{sec:ordinary-encoding}

Let $g\ge2$, $m\ge1$, and let $V$ be a finite universe.
Consider the database
\[
 S=\mathop{\biguplus}_{i=1}^g\{(i,x):x\in X_i\},
 \qquad |X_i|=m,\qquad N=gm.
\]
For sign assignments $p_i:V\to\{-1,1\}$, denote
\[
 \mu_i=\frac1m\sum_{x\in X_i}p_i(x),\qquad
 \mu=(\mu_1,\ldots,\mu_g).
\]
We want to estimate $\mu$ using subset queries on $S$.
Define
\[
 G_m=\{-1+2r/m:r=0,\ldots,m\}\cup\{0\},\qquad
 \mathcal F=G_m^g.
\]
Thus, $\mu\in\mathcal F\subseteq[-1,1]^g$, and $0\in\mathcal F$.

The mechanism of Theorem~\ref{thm:querylower} answers only subset queries on the whole tagged database $S$, and each answer is normalized by $N=gm$ rather than by the group size $m$.
The mean $\mu_i$ of a single group can be read off from one query, but at a high price: the query $\{(i,x):p_i(x)=1\}$ has true answer $(1+\mu_i)/(2g)$, so an answer within $\alpha$ determines $\mu_i$ only within $2g\alpha$, and asking for the $g$ means one group at a time gives an $\ell_1$ error of $2g^2\alpha$. The encoding below instead mixes all groups in every query, and recovers $\mu$ from a sequence of $M$ such queries with $\ell_1$ error $O(g\sqrt g\,\alpha)$.

\begin{lemma}[Robust encoding by subset queries]
\label{lem:robust-boolean-encoding}
There is a public deterministic encoding of each vector of $g$ sign
assignments into $M=O(g\log(m+2))$ subset queries.
For every public error threshold $\alpha\ge0$, there is a public
decoder with output $Y\in\mathcal F$ such that the following holds.
For every fixed database of the above form and every fixed full list
of queries, if each encoded answer has error at most $\alpha$
with probability at least $0.99$, then
\[
 \mathbb E\|Y-\mu\|_1\le16g\sqrt g\,\alpha+0.16g.
\]
The encoding and decoding preserve replacement differential privacy
and prefix forgetting at block boundaries, with the same parameters.
\end{lemma}

\begin{proof}
We first choose the public matrix used by the encoding.
Let $M$ be divisible by $32$. We claim that there is a matrix
$H\in\{-1,1\}^{M\times g}$ such that, for every nonzero
$v\in G_m^g-G_m^g$,
\begin{equation}
 \left|\left\{\ell\in[M]:
 |\langle H_{\ell,*},v\rangle|\ge\tfrac14\|v\|_2
 \right\}\right|\ge9M/32.
 \label{eq:ordinary-sign-separation}
\end{equation}
To see this, choose the rows independently and uniformly from
$\{-1,1\}^g$. 

Fix $v\ne0$ and write $a_i=v_i/\|v\|_2$ and
$Z=\langle h,v\rangle/\|v\|_2$ for one row $h$. Then 

\[
 \E Z^2=\sum_i a_i^2\,\E h_i^2+\sum_{i\ne k}a_ia_k\,\E[h_ih_k]=\sum_i a_i^2=1,
\]
since $h_i^2=1$ and $\E[h_ih_k]=\E h_i\,\E h_k=0$ for $i\ne k$.

To bound the fourth moment, note that every monomial in which some $h_i$ appears with an odd power has mean zero, so only two kinds of monomials survive: one index repeated four times, and a pair of distinct indices each repeated twice; the latter can be ordered in $\binom42=6$ ways. Therefore
\[
 \E Z^4=\sum_ia_i^4+6\sum_{i<k}a_i^2a_k^2
 =3\Bigl(\sum_ia_i^2\Bigr)^2-2\sum_ia_i^4
 =3-2\sum_ia_i^4\le3,
\]
using $(\sum_ia_i^2)^2=\sum_ia_i^4+2\sum_{i<k}a_i^2a_k^2$.

With $E=\{|Z|\ge1/4\}$ and $p=\Prb(E)$, $Z^2<\tfrac1{16}$ on $E^c$, and Cauchy--Schwarz with $X=Z^2$, $Y=\one_E$,
\[
 1=\E Z^2=\E[Z^2\one_{E^c}]+\E[Z^2\one_E]
 \le\frac1{16}+\sqrt{\E Z^4}\sqrt{\E\one_E^2}
 \le\frac1{16}+\sqrt{3p},
 \qquad\text{hence}\qquad
 p\ge\frac9{32}+\frac3{256}.
\]

For a fixed $v$ let $C_v$ be the number of rows of $H$ for which $|Z|\ge\tfrac14$. It is a sum of $M$ independent Bernoulli variables with $\E C_v=pM$. Hoeffding's inequality gives
\[
 \Prb\Bigl[C_v<\frac{9M}{32}\Bigr]
 \le\Prb\Bigl[\frac{C_v}{M}-p<-\frac3{256}\Bigr]
 \le\exp\Bigl(-2M\Bigl(\frac3{256}\Bigr)^2\Bigr)=e^{-9M/32768}.
\]
The grid $G_m$ has $m+2$ points, so there are at most $(m+2)^{2g}$ differences. 
By the union bound, the random matrix fails \eqref{eq:ordinary-sign-separation} for some difference with probability below $1$ for $M=O(g\log(m+2))$ with a sufficiently large constant. A random matrix therefore satisfies \eqref{eq:ordinary-sign-separation} for every difference simultaneously with positive probability, so a suitable deterministic $H$ exists.

Fix such a matrix $H$ and a public ordering of $\mathcal F$. Our encoding and decoding work as follows:
\begin{enumerate}
    \item Given $(p_1,\ldots,p_g)$, ask the subset queries
    \begin{equation}
     Q_\ell=\{(i,x)\in[g]\times V:H_{\ell i}p_i(x)=1\},
     \qquad \ell\in[M].
     \label{eq:ordinary-boolean-encoding}
    \end{equation}
    Their normalized true answers are
    $|S\cap Q_\ell|/N=1/2+(H\mu)_\ell/(2g)$.
        \item Let $A_\ell$ denote the mechanism's answer to $Q_\ell$. Clip the answers $A_\ell$ to $[0,1]$ and set
    $z_\ell=g(2A_\ell-1)$.
    \item Return the first $y\in\mathcal F$ satisfying
    \begin{equation}
     \left|\{\ell\in[M]:|z_\ell-(Hy)_\ell|\le2g\alpha\}\right|
     \ge7M/8.
     \label{eq:ordinary-grid-decoder}
    \end{equation}
    If no such $y$ exists, return zero.
\end{enumerate}
\paragraph{Privacy and prefix forgetting.}
Replacing one row of some $X_i$ replaces one row of $S$, and the
decoded answers are postprocessing of the scalar answers.
To erase a prefix of blocks, replace all queries in those blocks
by dummies. Applying prefix forgetting at that boundary and then
decoding the suffix preserves the same $(\rho,\eta)$ parameters,
since each block is encoded in advance, and its decoder uses only that
block's answers.

\paragraph{Utility.}
Since each group has exactly $m$ rows and $\one\{H_{\ell i}p_i(x)=1\}=\tfrac12(1+H_{\ell i}p_i(x))$,
\[
 \frac{|S\cap Q_\ell|}{N}
 =\frac1{gm}\sum_{i=1}^g\sum_{x\in X_i}\frac{1+H_{\ell i}p_i(x)}2
 =\frac12+\frac1{2gm}\sum_{i=1}^gH_{\ell i}\cdot m\mu_i
 =\frac12+\frac{(H\mu)_\ell}{2g}.
\]
Consequently, an answer with error at most $\alpha$ satisfies $|z_\ell-(H\mu)_\ell|=2g\,\bigl|A_\ell-|S\cap Q_\ell|/N\bigr|\le2g\alpha$; clipping to $[0,1]$ can only move $A_\ell$ closer to the true answer, which lies in $[0,1]$.

Assume first that at most $M/8$ answers have error greater than
$\alpha$. Clipping does not increase the error, so $\mu$ satisfies
\eqref{eq:ordinary-grid-decoder}. The decoder therefore returns a
feasible $Y$, so each of $|z_\ell-(HY)_\ell|\le2g\alpha$ and $|z_\ell-(H\mu)_\ell|\le2g\alpha$ holds on at least $7M/8$ rows, and both hold together on at least $3M/4$ rows (even when $Y\ne\mu$, in which case the two sets of rows need not coincide). On every such row the triangle inequality gives
\[
 |(H(Y-\mu))_\ell|=\bigl|(z_\ell-(H\mu)_\ell)-(z_\ell-(HY)_\ell)\bigr|
 \le|z_\ell-(H\mu)_\ell|+|z_\ell-(HY)_\ell|\le4g\alpha .
\]
If $Y\ne\mu$, the difference $v=Y-\mu$ is a nonzero element of $G_m^g-G_m^g$, since $Y,\mu\in\mathcal F$, so \eqref{eq:ordinary-sign-separation} applies to it: at least $9M/32$ rows satisfy $|(H(Y-\mu))_\ell|\ge\tfrac14\|Y-\mu\|_2$. At most $M/4=8M/32$ rows violate the bound $|(H(Y-\mu))_\ell|\le4g\alpha$, so some row satisfies both bounds, giving $\tfrac14\|Y-\mu\|_2\le4g\alpha$; this holds trivially also when $Y=\mu$. Multiplying by $4$, and then using $\|Y-\mu\|_1\le\sqrt g\,\|Y-\mu\|_2$, yields
\begin{equation}
 \|Y-\mu\|_2\le16g\alpha,\qquad
 \|Y-\mu\|_1\le16g\sqrt g\,\alpha.
 \label{eq:ordinary-good-decoding}
\end{equation}
It remains to handle the case in which more than $M/8$ answers have error greater than $\alpha$; we bound its probability and combine the two cases. Fix the database and the full list of queries.
Let $b$ be the number of answers in the block whose error relative to the true answer $|S\cap Q_\ell|/N$ exceeds $\alpha$.
Then $\E b\le0.01M$ by linearity of expectation, even if the answers are correlated (no independence or concentration is needed), and Markov's inequality gives $\Prb[b>M/8]\le0.01M/(M/8)=0.08$. On the complementary event \eqref{eq:ordinary-good-decoding} holds; on the event $\{b>M/8\}$, which includes the case in which the decoder finds no feasible $y$ and returns $0$, $\|Y-\mu\|_1\le2g$ since $Y,\mu\in[-1,1]^g$. Splitting the expectation according to this event gives
\begin{align}
 \E_{\rm coins}\|Y-\mu\|_1
 \le16g\sqrt g\,\alpha\cdot\Prb[b\le M/8]+2g\cdot\Prb[b>M/8]
 \le16g\sqrt g\,\alpha+0.16g .
 \label{eq:ordinary-expected-decoding}
\end{align}
\end{proof}

\subsection{The hard input distribution}
\label{sec:ordinary-source}

Fix a sufficiently large $m$ divisible by $40$, and put
\[
 \kappa=m/20,\qquad \ell=(m-\kappa)/2.
\]
Fix also a number of groups $g\ge2$, a universe $V$ of size $u>m$,
and a number of blocks $K$. Our hard database has $N=gm$ records
over $[g]\times V$, with $m$ records of each tag. In every group,
$\kappa$ rows are hidden and the remaining $2\ell$ rows form the core.
The mechanism is still defined on every $N$-row database.
Suppose it satisfies $(\eps,\delta)$-DP, $(\rho,\eta)$-prefix
forgetting, and accuracy $(\alpha,0.01)$.
We encode each block using Lemma~\ref{lem:robust-boolean-encoding}
and decode its $M=O(g\log(m+2))$ answers into $Y_j\in[-1,1]^g$.

Independently for each group $i\in[g]$:
\begin{enumerate}
    \item Choose $X_i$ uniformly among ordered distinct $m$-tuples
    from $V$. Choose a uniform $\kappa$-subset $J_i$ of its slots.
    Denote the values in these hidden slots by
    $\xi_{i1},\ldots,\xi_{i\kappa}$, and the remaining ordered rows
    by $W_i$. 
    \item Independently sample $Z_i\in\{-\ell,\ldots,\ell\}$ with
    \begin{equation*}
     r=\frac{m-3}{m+3},\qquad
     \Prb[Z_i=z]=c_\ell r^{|z|},\qquad
     c_\ell=\left(1+2\sum_{z=1}^{\ell}r^z\right)^{-1}.
    \end{equation*}
    Choose a uniform $(\ell+Z_i)$-subset of $W_i$ to receive sign
    $+1$, and give the other core rows sign $-1$. These signs are
    fixed in every block.
    \item For every $x\in V_i:=V\setminus W_i$ and $j\in[K]$,
    choose $p_{ij}(x)$ independently and uniformly from $\{-1,1\}$.
\end{enumerate}

The database is
\[
 S=\biguplus_{i=1}^g\{(i,x):x\in X_i\},\qquad X=(X_1,\ldots,X_g),
\]
and block $j$ encodes $(p_{1j},\ldots,p_{gj})$
using~\eqref{eq:ordinary-boolean-encoding}. Thus the database, the hidden slots and the core signs are common to all blocks; only the signs on $V_i$, which include those of the hidden rows, change from block to block. Write
\[
 B_{ij}=\sum_{a=1}^{\kappa}p_{ij}(\xi_{ia}),\qquad
 t_{ij}=\sum_{x\in X_i}p_{ij}(x)=\,2Z_i+B_{ij},\qquad \mu_{ij}=t_{ij}/m.
\]
For every fixed block, the hidden signs are independent fair signs
and are independent of the $Z_i$. In particular,
\[
 \E B_{ij}=0,\qquad \E B_{ij}^2=\E\Bigl(\sum_{a=1}^\kappa p_{ij}(\xi_{ia})\Bigr)^2
 =\sum_a\E\bigl[p_{ij}(\xi_{ia})^2\bigr]+\sum_{a\ne b}\E p_{ij}(\xi_{ia})\,\E p_{ij}(\xi_{ib})=\kappa .
\]
Therefore
\begin{equation}
 \E[B_{ij}\mu_{ij}]=\frac{\E[B_{ij}(2Z_i+B_{ij})]}m
 =\frac{2\,\E Z_i\,\E B_{ij}+\E B_{ij}^2}m=\frac\kappa m, \qquad \E\sum_i B_{ij}\mu_{ij}=\frac{g\kappa}{m}.
  \label{eq:ordinary-core-true-score}
\end{equation}
The quantity $g\kappa/m$ in \eqref{eq:ordinary-core-true-score} is the correlation of the \emph{true} normalized answers with the hidden signs, and it does not depend on the distribution of $Z_i$. However, the proof below measures how much the mechanism's answers correlate with the signs that the current block assigns to these particular rows. Restricted to the $m$ rows of $X_i$, the sign assignments $p_{ij}$ of the different blocks agree on the $2\ell$ core rows and are drawn afresh on the $\kappa$ hidden rows. Within a single block the hidden slots cannot be identified: the set $J_i$ is a label of the analysis and not an input of the mechanism, and the database looks the same whichever slots are hidden (however, the block's signs do provide some statistical evidence, though only a weak one: given the database and the block's signs, each hidden sign is biased toward the majority sign by only $3/m$, as quantified in Lemma~\ref{lem:ordinary-core-posterior}). A mechanism that remembers earlier blocks, however, can identify them: across blocks they are exactly the slots whose signs ``flicker''. This is what prefix forgetting takes away.

\begin{lemma}[Posterior of the hidden signs]
\label{lem:ordinary-core-posterior}
Fix $j$, let $\mathcal C=(X,p_{\cdot j})$, and put
$v_i=\E[B_{ij}\mid\mathcal C]$.
The variables $B_{ij}$ are conditionally independent across groups,
and, writing $(z)_+=\max\{z,0\}$,
\begin{equation*}
 \E\left(|v_i|-\frac{3\kappa}{m}\right)_+
 \le\frac{\kappa}{5m}.
\end{equation*}
\end{lemma}

\begin{proof}
Fix the block $j$ and a group $i$. Mark a uniformly random hidden slot of group $i$ and let $b$ be
its current sign. Condition on $\mathcal C=(X,p_{\cdot j})$ and on $\mathcal H$---the identities of the other $\kappa-1$ hidden slots together with their current signs. Given $X$ and $p_{ij}$, the sign $p_{ij}(X_i[k])$ of every slot $k$ is known. The values of $p_{ij}$ outside $X_i$ are also observed, but they are independent of all hidden variables and contribute a common factor to every likelihood below. Let $T\subseteq[m]$ be the set of the $2\ell+1$ slots not revealed by $\mathcal H$ (the core slots and the marked slot). What remains hidden is the position of the marked slot among the $2\ell+1$ slots of $T$ (nothing in $X_i$ or in the signs identifies it), the value $Z_i$, the set $W_i^+\subseteq W_i$ of core rows that received sign $+1$ (of size $\ell+Z_i$), and the sign $b$ of the marked slot. Let
\[
 T^+=\{k\in T:\ p_{ij}(X_i[k])=+1\},\qquad {n_+}=|T^+|=\frac{2\ell+1+{\tilde t}}2,\qquad {n_-}=2\ell+1-{n_+},
\]
where ${\tilde t}=\sum_{k\in T}p_{ij}(X_i[k])$ is the sign sum over the slots of $T$ and ${n_-}$ is the number of its negative slots. The generative process gives
\[
 T^+=W_i^+\cup\{\text{marked slot}\}\ \text{ if }b=+1,\qquad
 T^+=W_i^+\ \text{ if }b=-1 ,
\]
so $b=+1$ exactly when the marked slot is one of the ${n_+}$ slots of $T^+$ and $b=-1$ exactly when it is one of the other ${n_-}$ slots. Hence $b$ is random given the observation $(\mathcal C,\mathcal H)$, and $z=\E[b\mid\mathcal C,\mathcal H]$ is its mean under the posterior of the marked slot and of $Z_i$.

\paragraph{Computing the posterior of $b$.}
Write $\mathcal G$ for the information in $(\mathcal C,\mathcal H)$ other than the signs on $T$: the database $X$, the identities and signs of the other $\kappa-1$ hidden slots, the values of $p_{ij}$ outside $X_i$, and the data of the other groups. The observation $T^+$ and the hypothesis on $b$ are what remains, and the posterior of $b$ is determined by the ratio of the two joint probabilities $\Prb[T^+,\ b=+1\mid\mathcal G]$ and $\Prb[T^+,\ b=-1\mid\mathcal G]$.

\paragraph{Interior case:} Let $T^+$ be the observed set, and consider first the interior case $0<{n_+}<2\ell+1$. If $b=+1$, three things are required: $|W_i^+|={n_+}-1$, i.e.\ $Z_i={n_+}-1-\ell=({\tilde t}-1)/2$; the marked slot belongs to $T^+$; and $W_i^+=T^+\setminus\{\text{marked slot}\}$. Summing over the position $k$ of the marked slot, whose prior is uniform on the $2\ell+1$ slots (all factors below are conditional on $\mathcal G$; it is omitted from the notation because $Z_i$, $b$ and the position of the marked slot in $T$ do not depend on it),
\begin{align*}
 \MoveEqLeft\Prb[T^+,\ b=+1\mid\mathcal G]\\
 &=\Prb\Bigl[Z_i=\tfrac{{\tilde t}-1}2\Bigr]\cdot\frac12\cdot\sum_{k \in T}\frac1{2\ell+1}\cdot
   \Prb\bigl[W_i^+=T^+\setminus\{k\}\ \big|\ \text{marked}=k,\ |W_i^+|={n_+}-1\bigr]\\
 &=\Prb\Bigl[Z_i=\tfrac{{\tilde t}-1}2\Bigr]\cdot\frac12\cdot\sum_{k\in T^+}\frac1{2\ell+1}\cdot\frac1{\binom{2\ell}{{n_+}-1}}
  =\Prb\Bigl[Z_i=\tfrac{{\tilde t}-1}2\Bigr]\cdot\frac12\cdot\frac{{n_+}}{(2\ell+1)\binom{2\ell}{{n_+}-1}},
\end{align*}
where the second equality holds because for $k\notin T^+$ the summand vanishes: $T^+\setminus\{k\}=T^+$ has size ${n_+}$ while the conditioning fixes $|W_i^+|={n_+}-1$, so the event is empty. The same computation gives
\begin{align*}
 \MoveEqLeft\Prb[T^+,\ b=-1\mid\mathcal G]\\
 &=\Prb\Bigl[Z_i=\tfrac{{\tilde t}+1}2\Bigr]\cdot\frac12\cdot\sum_{k\in T}\frac1{2\ell+1}\cdot
   \Prb\bigl[W_i^+=T^+\ \big|\ \text{marked}=k,\ |W_i^+|={n_+}\bigr]\\
 &=\Prb\Bigl[Z_i=\tfrac{{\tilde t}+1}2\Bigr]\cdot\frac12\cdot\sum_{k\notin T^+}\frac1{2\ell+1}\cdot\frac1{\binom{2\ell}{{n_+}}}
  =\Prb\Bigl[Z_i=\tfrac{{\tilde t}+1}2\Bigr]\cdot\frac12\cdot\frac{{n_-}}{(2\ell+1)\binom{2\ell}{{n_+}}}.
\end{align*}
The two combinatorial factors coincide:
\begin{align*}
 \frac{{n_+}}{(2\ell+1)\binom{2\ell}{{n_+}-1}}
 &=\frac{{n_+}\,({n_+}-1)!\,{n_-}!}{(2\ell+1)!}
 =\frac1{\binom{2\ell+1}{n_+}},\\
 \frac{{n_-}}{(2\ell+1)\binom{2\ell}{{n_+}}}
 &=\frac{{n_-}\,{n_+}!\,({n_-}-1)!}{(2\ell+1)!}
 =\frac1{\binom{2\ell+1}{n_+}}.
\end{align*}

In words: for every specific set $T^+$ of size ${n_+}$, the probability of observing exactly $T^+$ is the same, $1/\binom{2\ell+1}{n_+}$, under both $b=+1$ and $b=-1$. The posterior of $b$ given $\mathcal C,\mathcal H$ is nevertheless slightly biased, because the two hypotheses require different values of $Z_i$, namely $({\tilde t}-1)/2$ and $({\tilde t}+1)/2$, and the prior masses of $Z_i$ at these two values differ (by the factor $r$), as follows.
Conditioning on $(\mathcal C,\mathcal H)$ is the same as conditioning on $(T^+,\mathcal G)$, so by Bayes' rule and the two joint probabilities computed above,
\begin{align*}
 \Prb[b=+1\mid\mathcal C,\mathcal H]
 &=\frac{\Prb[T^+,\ b=+1\mid\mathcal G]}{\Prb[T^+,\ b=+1\mid\mathcal G]+\Prb[T^+,\ b=-1\mid\mathcal G]}\\
 &=\frac{\Prb\bigl[Z_i=\frac{{\tilde t}-1}2\bigr]\cdot\frac12\cdot\frac1{\binom{2\ell+1}{n_+}}}
        {\Prb\bigl[Z_i=\frac{{\tilde t}-1}2\bigr]\cdot\frac12\cdot\frac1{\binom{2\ell+1}{n_+}}
        +\Prb\bigl[Z_i=\frac{{\tilde t}+1}2\bigr]\cdot\frac12\cdot\frac1{\binom{2\ell+1}{n_+}}}
 =\frac{\Prb\bigl[Z_i=\frac{{\tilde t}-1}2\bigr]}{\Prb\bigl[Z_i=\frac{{\tilde t}-1}2\bigr]+\Prb\bigl[Z_i=\frac{{\tilde t}+1}2\bigr]} .
\end{align*}
The complementary probability $\Prb[b=-1\mid\mathcal C,\mathcal H]$ is the same fraction with $\Prb\bigl[Z_i=\frac{{\tilde t}+1}2\bigr]$ in the numerator. The number ${\tilde t}$ is a sum of $2\ell+1$ signs, hence odd and nonzero, and
\[
 \Bigl|\frac{{\tilde t}-\sgn({\tilde t})}2\Bigr|=\frac{|{\tilde t}|-1}2,\qquad
 \Bigl|\frac{{\tilde t}+\sgn({\tilde t})}2\Bigr|=\frac{|{\tilde t}|+1}2 .
\]
When $|{\tilde t}|<2\ell+1$, oddness gives $|{\tilde t}|\le2\ell-1$, so both values are at most $\ell$ and lie in the support $\{-\ell,\ldots,\ell\}$ of $Z_i$, the prior masses are $c_\ell r^{(|{\tilde t}|-1)/2}$ and $c_\ell r^{(|{\tilde t}|+1)/2}$, and therefore,
\begin{gather*}
 \Prb[b=\sgn({\tilde t})\mid\mathcal C,\mathcal H]
 =\frac{c_\ell r^{(|{\tilde t}|-1)/2}}{c_\ell r^{(|{\tilde t}|-1)/2}+r\,c_\ell r^{(|{\tilde t}|-1)/2}}
 =\frac1{1+r},\qquad
 \Prb[b=-\sgn({\tilde t})\mid\mathcal C,\mathcal H]=\frac r{1+r},\\
 \begin{aligned}
 z=\E[b\mid\mathcal C,\mathcal H]
 &=\sgn({\tilde t})\Bigl(\frac1{1+r}-\frac r{1+r}\Bigr)
 =\frac{1-r}{1+r}\sgn({\tilde t})\\
 &=\frac{6/(m+3)}{2m/(m+3)}\sgn({\tilde t})
 =\frac3m\sgn({\tilde t})=\theta\sgn({\tilde t}),
 \end{aligned}
\end{gather*}
using $1-r=\frac6{m+3}$ and $1+r=\frac{2m}{m+3}$ for $r=\frac{m-3}{m+3}$.

\paragraph{Endpoint case:} When $|{\tilde t}|=2\ell+1$ all $2\ell+1$ signs are equal, so the marked slot has this sign whichever slot it is; hence $z=\sgn({\tilde t})$ and $|z|=1$.

Altogether, $z=\E[b\mid\mathcal C,\mathcal H]$ is given in both cases by
\[
 z=\begin{cases}
 \theta\,\sgn({\tilde t}), & |{\tilde t}|<2\ell+1,\\[2pt]
 \sgn({\tilde t}), & |{\tilde t}|=2\ell+1.
 \end{cases}
\]

\paragraph{The endpoint probability $p_{\rm end}$.} The event ${\tilde t}=2\ell+1$ requires $Z_i=\ell$ (all core rows positive) and $b=+1$; the event ${\tilde t}=-(2\ell+1)$ requires $Z_i=-\ell$ and $b=-1$. Since $b$ is a fair sign independent of $Z_i$ and $\Prb[Z_i=\pm\ell]=c_\ell r^\ell$,
\begin{gather*}
 c_\ell^{-1}=1+2\sum_{z=1}^\ell r^z=1+\frac{2r(1-r^\ell)}{1-r}=\frac{1+r-2r^{\ell+1}}{1-r},\\
 p_{\rm end}=2\cdot c_\ell r^\ell\cdot\frac12 =c_\ell r^\ell=\frac{(1-r)r^\ell}{1+r-2r^{\ell+1}} .
\end{gather*}
Here $m(1-r)=\frac{6m}{m+3}\to6$, and since $\ell=\frac{m-\kappa}2=\frac{19m}{40}$,
\begin{gather*}
 r^\ell=\Bigl(1-\frac6{m+3}\Bigr)^{19m/40}\longrightarrow e^{-6\cdot19/40}=e^{-57/20},\\
 m\,p_{\rm end}\longrightarrow\frac{6e^{-57/20}}{2-2e^{-57/20}}=\frac3{e^{57/20}-1}\approx0.184<\frac15 .
\end{gather*}

\paragraph{From the marked slot to $v_i$.} By definition $v_i=\E[B_{ij}\mid\mathcal C]=\sum_{a=1}^\kappa\E[p_{ij}(\xi_{ia})\mid\mathcal C]$. The marked slot is uniform among the $\kappa$ hidden slots and independent of everything else, so $\E[b\mid\mathcal C]=\frac1\kappa\sum_a\E[p_{ij}(\xi_{ia})\mid\mathcal C]=v_i/\kappa$. By the tower property, $\E[b\mid\mathcal C]=\E\bigl[\E[b\mid\mathcal C,\mathcal H]\mid\mathcal C\bigr]=\E[z\mid\mathcal C]$. The function $x\mapsto(|x|-\theta)_+$ is convex, so Jensen's inequality gives, conditionally on $\mathcal C$,
\[
 \bigl(|v_i|-\kappa\theta\bigr)_+=\kappa\bigl(|\E[z\mid\mathcal C]|-\theta\bigr)_+
 \le\kappa\,\E\bigl[(|z|-\theta)_+\ \big|\ \mathcal C\bigr].
\]
Now $(|z|-\theta)_+=0$ in the interior case and $(|z|-\theta)_+=1-\theta\le1$ at an endpoint, so $\E(|z|-\theta)_+\le\Prb[|{\tilde t}|=2\ell+1]=p_{\rm end}\le1/(5m)$, which proves the lemma's inequality. The quantity $3\kappa/m=\kappa\theta$ in the lemma is thus the interior bias $\theta=3/m$ per hidden slot, accumulated over the $\kappa$ hidden slots of the group.
\paragraph{Conditional independence across groups.} Since all the sampling of group $i$ ($X_i$, $J_i$, $Z_i$, $W_i^+$ and the signs of $p_{ij}$ outside $X_i$) is independent across groups, the joint law of $(X,p_{\cdot j},B_{1j},\ldots,B_{gj})$ is a product over $i$, and so is the conditional law of $(B_{1j},\ldots,B_{gj})$ given $\mathcal C$; hence the $B_{ij}$ are conditionally independent, each depending on its own group's data only.
\end{proof}

\subsection{Accuracy and prefix forgetting correlate the answers with the hidden signs}
\label{sec:ordinary-score}

Fix a block $j$ and denote
$\mu_j=(\mu_{1j},\ldots,\mu_{gj})$ and
$B_j=(B_{1j},\ldots,B_{gj})$.
Let $P$ denote the original execution. Since $j$ is fixed, the index $j$ is omitted from the vector $v=(v_1,\ldots,v_g)$ of Lemma~\ref{lem:ordinary-core-posterior}. Note that $v$ depends on block $j$ only, through $p_{\cdot j}$, and on no other block.

\paragraph{Using accuracy.} Since $\mu_{ij}=t_{ij}/m$ is a function of $\mathcal C$ only, it is a constant once $\mathcal C$ is given, and can therefore be moved inside the conditional expectation defining $v_i$: $v_i\mu_{ij}=\E[B_{ij}\mid\mathcal C]\,\mu_{ij}=\E[B_{ij}\mu_{ij}\mid\mathcal C]$. Taking expectations, summing over $i$ and using the tower property and \eqref{eq:ordinary-core-true-score} gives
\[
 \E\langle v,\mu_j\rangle=\sum_{i=1}^g\E\bigl[\E[B_{ij}\mu_{ij}\mid\mathcal C]\bigr]=\sum_{i=1}^g\E[B_{ij}\mu_{ij}]=\frac{g\kappa}m.
\]

For every realization,
\begin{gather*}
 \langle v,Y_j\rangle=\langle v,\mu_j\rangle+\sum_{i=1}^gv_i\,(Y_{ij}-\mu_{ij}),\\
 v_i(Y_{ij}-\mu_{ij})\ \ge\ -|v_i|\,|Y_{ij}-\mu_{ij}|
 \ \ge\ -\frac{3\kappa}m|Y_{ij}-\mu_{ij}|-2\Bigl(|v_i|-\frac{3\kappa}m\Bigr)_+ ,
\end{gather*}
using $|v_i|\le\frac{3\kappa}m+(|v_i|-\frac{3\kappa}m)_+$ and $|Y_{ij}-\mu_{ij}|\le2$. Taking expectations and using Lemma~\ref{lem:ordinary-core-posterior} for the last term,
\[
 \E_P\langle v,Y_j\rangle
 \ \ge\ \underbrace{\E\langle v,\mu_j\rangle}_{=\,g\kappa/m}
 -\frac{3\kappa}m\E_P\|Y_j-\mu_j\|_1
 -2\sum_{i=1}^g\underbrace{\E\Bigl(|v_i|-\frac{3\kappa}m\Bigr)_+}_{\le\,\kappa/(5m)} .
\]

Inserting \eqref{eq:ordinary-expected-decoding}, which holds for each fixed sampled input and hence also after averaging over the input,
\begin{align}
 \E_P\langle v,Y_j\rangle 
 &\ge \frac{g\kappa}m-\frac{3\kappa}m\bigl(16g\sqrt g\,\alpha+0.16g\bigr)-\frac{2g\kappa}{5m}\notag\\
 &= \frac{g\kappa}m -\Bigl(\frac{0.48g\kappa}m +\frac{48 g\sqrt g\kappa}m\,\alpha\Bigr)-\frac{2g\kappa}{5m}\notag\\
 &\ge \frac{g\kappa}m\Bigl(\frac1{10}-48\sqrt g\,\alpha\Bigr).
 \label{eq:ordinary-core-boundary-score}
\end{align}

\begin{remark}[The role of the prior of $Z_i$ in \eqref{eq:ordinary-core-boundary-score}]
\label{rem:prior-choice}
The two loss terms in \eqref{eq:ordinary-core-boundary-score}, $\frac{3\kappa}m\E_P\|Y_j-\mu_j\|_1$ and $2\sum_i\E(|v_i|-\frac{3\kappa}m)_+$, are where the choice of the geometric prior enters, and the reason for this choice is the following. The unconditional correlation $\E\langle B_j,\mu_j\rangle=g\kappa/m$ is the same for every distribution of $Z_i$: its computation uses only that the hidden signs are fair and independent of $Z_i$. What the distribution of $Z_i$ determines is how this total is split across the groups once one conditions on the current block: the prior of $Z_i$ decides whether the summands of $\sum_i\E[v_i\mu_{ij}]$ are concentrated on a small fraction of the groups (those that contain endpoint slots), or spread as a small bias over all groups.

Under a uniform prior on $\{-\ell,\ldots,\ell\}$ the two probabilities $\Prb\bigl[Z_i=\tfrac{{\tilde t}\mp1}2\bigr]$ compared in the computation of the posterior are equal, so in the interior case the hidden sign is exactly fair and its conditional mean vanishes; only the endpoint case contributes:
\[
 z=\sgn(\tilde t)\,\one\{|\tilde t|=2\ell+1\},
 \qquad
 \Prb\bigl[|\tilde t|=2\ell+1\bigr]=\frac1{2\ell+1}>\frac1m .
\]
A group therefore contains about $\kappa/m$ endpoint slots in expectation, so the endpoints contribute about $\kappa/m$ to $\E|v_i|$, five times the bound $\kappa/(5m)$ that the geometric prior gives for the corresponding term. The decomposition used in \eqref{eq:ordinary-core-boundary-score} then gives nothing: the interior bias is $0$ instead of $3\kappa/m$, so its last term becomes $2\sum_i\E|v_i|$, and since $|\mu_{ij}|\le1$,
$2\sum_{i=1}^g\E|v_i|\ \ge\ 2\,\E\langle v,\mu_j\rangle=\frac{2g\kappa}m$, at least twice the signal. 
So we need to use the trivial decomposition $\langle v,Y_j\rangle=\langle v,\mu_j\rangle+\sum_iv_i(Y_{ij}-\mu_{ij})$ and bound its loss term directly. Under the uniform prior $v_i$ vanishes outside a fraction of about $\kappa/m=1/20$ of the groups, and the only control on $Y_j-\mu_j$ is \eqref{eq:ordinary-expected-decoding}, which bounds $\E\|Y_j-\mu_j\|_1$ by about $0.16g$ but says nothing about how the error is distributed over the groups. This budget suffices for $|Y_{ij}-\mu_{ij}|\approx2$ on every group with $v_i\ne0$, and then
\[
 \Bigl|\sum_iv_i(Y_{ij}-\mu_{ij})\Bigr|\approx2\sum_i|v_i|,
 \qquad
 2\sum_i\E|v_i|\approx\frac{2g\kappa}m ,
\]
about twice the signal, so no positive lower bound on $\E_P\langle v,Y_j\rangle$ follows.

With the geometric prior the picture is different. In the interior groups the bias is not zero, but it is small: $|v_i|\le3\kappa/m$. Larger values of $|v_i|$ arise only from endpoint slots, and these are rare: each hidden slot is at an endpoint with probability $p_{\rm end}\le1/(5m)$, and this rarity is a consequence of the choice $\Prb[Z_i=z]\propto r^{|z|}$. Consequently, even with the crudest bound $|Y_{ij}-\mu_{ij}|\le2$, the endpoint term is small:
$2\sum_{i=1}^g\E\Bigl(|v_i|-\frac{3\kappa}m\Bigr)_+\le0.4\cdot\frac{g\kappa}m$, well below the signal $g\kappa/m$, which therefore survives.

The constant $3$ in $r=(m-3)/(m+3)$ balances the two loss terms of \eqref{eq:ordinary-core-boundary-score}, which, apart from the term with $\alpha$, are
\[
 L_1=\frac{3\kappa}m\,\E_P\|Y_j-\mu_j\|_1\le0.48\,\frac{g\kappa}m
 \qquad\text{and}\qquad
 L_2=2\sum_i\E\Bigl(|v_i|-\frac{3\kappa}m\Bigr)_+\le0.4\,\frac{g\kappa}m.
\]
For a general $r=(m-c)/(m+c)$ the interior bias per hidden slot is $\frac{1-r}{1+r}=\frac cm$, so $L_1\le\frac{c\kappa}m\cdot0.16g=0.16c\,\frac{g\kappa}m$, while $m(1-r)\to2c$ and $r^\ell\to e^{-19c/20}$ give
\[
 m\,p_{\rm end}\longrightarrow\frac{2c\,e^{-19c/20}}{2-2e^{-19c/20}}=\frac{c}{e^{19c/20}-1},
 \qquad
 L_2\le2g\kappa\,p_{\rm end}\approx\frac{2c}{e^{19c/20}-1}\,\frac{g\kappa}m .
\]
In units of the signal $g\kappa/m$ the total loss is therefore
\begin{gather*}
 \Lambda(c)=0.16c+\frac{2c}{e^{19c/20}-1}:\\
 \Lambda(c)\xrightarrow[c\to0]{}\frac{40}{19}>1,\qquad
 \Lambda(3)=0.48+0.37=0.85,\qquad
 \Lambda(c)\xrightarrow[c\to\infty]{}\infty .
\end{gather*}
A small $c$ (the uniform prior in the limit) makes endpoints frequent and $L_2$ exceed the signal; a large $c$ makes the interior bias, and with it the price $L_1$ of the decoding error, too large. The choice $c=3$ keeps $1-\Lambda(3)=0.15>1/10$ of the signal.
\end{remark}

\paragraph{Using prefix forgetting.} The quantity needed in Section~\ref{sec:ordinary-soundness} is $\E_P\langle B_j,Y_j\rangle=\E_P\langle v,Y_j\rangle+\E_P\langle B_j-v,Y_j\rangle$. The first term was bounded in \eqref{eq:ordinary-core-boundary-score}; this paragraph bounds the second, $\E_P\psi_j(Y_j)$ for $\psi_j(y)=\langle B_j-v,y\rangle$ with $y\in[-1,1]^g$, which is where prefix forgetting enters.

Consider an execution $R_j$ in which all preceding blocks are replaced by dummies, and let $Y'_j$ be the decoded current answer. Given $\mathcal C=(X,p_{\cdot j})$, the remaining randomness consists of the mechanism's coins, the hidden slots $J_i$ (and with them $Z_i$ and $B_{ij}$), and the sign assignments $p_{\cdot j'}$ of the other blocks (which are drawn afresh outside the core, and although they agree with $p_{\cdot j}$ on the core rows, the locations of these rows are unknown). In $R_j$ the mechanism receives dummies in the earlier blocks and, by causality, does not see the later ones, so
\[
 Y'_j=f\bigl(X,\ p_{\cdot j},\ \text{block index},\ \text{coins}\bigr)
\]
for some function $f$. The coins are independent of the input, and the index set $J_i$ is not among the inputs: the hidden rows are ordinary rows of $X$, and nothing marks them. Since, given $\mathcal C$, the only remaining randomness in $B_{ij}=\sum_ap_{ij}(\xi_{ia})$ is through $J_i$, it follows that conditional on $\mathcal C$, $Y'_j$ is independent of $J_1,\ldots,J_g$ and therefore of $B_j$. In words, $Y'_j$ is certainly correlated with $B_j$, through their common dependence on $(X,p_{\cdot j})$, but it knows nothing about $B_j$ beyond what $\mathcal C$ already knows: $\Law(Y'_j\mid\mathcal C,B_j)=\Law(Y'_j\mid\mathcal C)$.

In the original execution this fails: $Y_j$ also depends on the earlier sign assignments $p_{\cdot j'}$, $j'<j$, which are not conditioned on and which are correlated with $J_i$. On the core rows the signs agree in every block, on the hidden rows they are redrawn, so comparing blocks reveals the slots whose signs flicker, that is, $J_i$; a mechanism that saw the earlier blocks can compute $J_i$ and output in block $j$ an answer that depends on $B_{ij}$, for instance $Y_{ij}=-\sgn(B_{ij})$. The conditional independence therefore holds only in $R_j$.

Since $Y'_j$ is independent of $B_j$ given $\mathcal C$, conditioning additionally on $Y'_j$ does not change the law of $B_j$: $\E[B_j\mid\mathcal C,Y'_j]=\E[B_j\mid\mathcal C]=v$, and the coordinates remain conditionally independent. Therefore,
\begin{align*}
 \E[\psi_j(Y'_j)\mid\mathcal C,Y'_j]&=\bigl\langle\E[B_j\mid\mathcal C,Y'_j]-v,\ Y'_j\bigr\rangle=\bigl\langle\E[B_j\mid\mathcal C]-v,\ Y'_j\bigr\rangle=0,\\
 \E\bigl[\psi_j(Y'_j)^2\mid\mathcal C,Y'_j\bigr]&=\operatorname{Var}(\psi_j(Y'_j)\mid\mathcal C,Y'_j)=\sum_{i=1}^g(Y'_{ij})^2\operatorname{Var}(B_{ij}\mid\mathcal C)\\
 &\le\sum_{i=1}^g\operatorname{Var}(B_{ij}\mid\mathcal C)\le\sum_{i=1}^g\E[B_{ij}^2\mid\mathcal C],
\end{align*}
where the variance of the sum splits by conditional independence and $|Y'_{ij}|\le1$ was used. Averaging over $(\mathcal C,Y'_j)$ and using \eqref{eq:ordinary-core-true-score} gives $\E_{R_j}\psi_j(Y'_j)^2\le\sum_i\E B_{ij}^2=g\kappa$, and Jensen's inequality gives 
 \[\E_{R_j}|\psi_j(Y'_j)|\le\sqrt{\E_{R_j}\psi_j(Y'_j)^2}\le\sqrt{g\kappa}, \qquad |\psi_j(y)|\le\sum_i|B_{ij}-v_i|\,|y_i|\le2g\kappa.
\]

If $P\approx_{a,b}Q$ and $f$ is measurable with $0\le f\le L$, then
\[
 \E_Pf=\int_0^L\Prb_P[f>s]\,ds\le\int_0^L\bigl(e^a\Prb_Q[f>s]+b\bigr)ds=e^a\E_Qf+Lb .
\]
For each fixed sampled input, prefix forgetting at the block boundary (followed by decoding, a postprocessing) makes the laws of $Y_j$ and $Y'_j$ $(\rho,\eta)$-close while $y\mapsto|\psi_j(y)|$ is a fixed function bounded by $2g\kappa$; applying the inequality above with $a=\rho$, $b=\eta$, $L=2g\kappa$ and averaging over the input gives
\begin{equation}
 \E_P|\psi_j(Y_j)|\le e^\rho\,\E_{R_j}|\psi_j(Y'_j)|+2g\kappa\eta\le e^\rho\sqrt{g\kappa}+2g\kappa\eta .
 \label{eq:ordinary-core-forgetting-transfer}
\end{equation}
Combining with~\eqref{eq:ordinary-core-boundary-score}, we obtain
\begin{equation}
 \E_P\langle B_j,Y_j\rangle=\E_P\langle v,Y_j\rangle+\E_P\psi_j(Y_j)\ge\E_P\langle v,Y_j\rangle-\E_P|\psi_j(Y_j)|
 \ge\frac{g\kappa}{m}
       \left(\frac1{10}-48\sqrt g\,\alpha\right)
       -e^\rho\sqrt{g\kappa}-2g\kappa\eta.
 \label{eq:ordinary-core-block-lower}
\end{equation}

For a group $i$ and a candidate record $x\in V_i$, denote
\[
 c_i(x)=\frac1K\sum_{j=1}^K p_{ij}(x)Y_{ij}.
\]
For every hidden record, $c_i(\xi_{ia})=\frac1K\sum_jp_{ij}(\xi_{ia})Y_{ij}$, so summing over $a$ and $i$ and exchanging the order of summation,
\[
 \sum_{i=1}^g\sum_{a=1}^\kappa c_i(\xi_{ia})
 =\frac1K\sum_{j=1}^K\sum_{i=1}^g\Bigl(\sum_{a=1}^\kappa p_{ij}(\xi_{ia})\Bigr)Y_{ij}
 =\frac1K\sum_{j=1}^K\sum_{i=1}^gB_{ij}Y_{ij}=\frac1K\sum_{j=1}^K\langle B_j,Y_j\rangle.
\]
Averaging~\eqref{eq:ordinary-core-block-lower} over $j$ and using
$(z)_+\ge z$ gives
\begin{equation}
 \E_P\frac1{g\kappa}\sum_{i,a}\bigl(c_i(\xi_{ia})\bigr)_+
 \ \ge\ \E_P\frac1{g\kappa}\sum_{i,a}c_i(\xi_{ia})
 \ \ge\ \frac1m\Bigl(\frac1{10}-48\sqrt g\,\alpha\Bigr)-\frac{e^\rho}{\sqrt{g\kappa}}-2\eta .
 \label{eq:add-scorelower}
\end{equation}

\subsection{Data privacy bounds the correlation}
\label{sec:ordinary-soundness}

We now use privacy to upper bound the same correlation.
We will need the following elementary fact about random sign matrices.

\begin{lemma}
\label{lem:random-sign-positive-parts}
Let $A\in\{-1,1\}^{v\times K}$ have independent fair entries.
For every random $y\in[-1,1]^K$, which may depend arbitrarily on
$A$ and additional randomness,
\begin{equation*}
 \E\left[\frac1v\sum_{w=1}^v
 \left(\frac1K\sum_{j=1}^K A_{wj}y_j\right)_+\right]
 \le\sqrt{2\log2\left(\frac1v+\frac1K\right)}.
\end{equation*}
\end{lemma}

\begin{remark}
The lemma will be applied as follows: the rows of $A$ are the sign vectors $(p_{ij}(x))_{j\in[K]}$ that the blocks assign to values $x$ that are not in the database of a group $i$, the vector $y$ consists of the mechanism's decoded answers $(Y_{ij})_j$, and we want the correlation $\frac1K\sum_jp_{ij}(x)Y_{ij}$ to be small. If the answers were independent of these signs, the correlation would be an average of $K$ independent fair signs with bounded coefficients: mean zero and standard deviation at most $1/\sqrt K$, hence of order $1/\sqrt K$ in a typical execution. However, even when the answers are not independent of the signs, the lemma quantifies what is nevertheless true: a single vector $y$ cannot be correlated with a sizeable fraction of many random sign vectors at once, even if it is chosen after seeing all of them. Precisely, the lemma takes the positive part of the correlation of $y$ with each row, averages it over the rows, and bounds the expectation of this average over the random signs. When the number of rows is at least $K$ the bound is of order $1/\sqrt K$, the same order as when $y$ is independent of the signs: arbitrary dependence costs only a constant factor.
\end{remark}

\begin{proof}
For every $A$ and $y$, the expression inside the expectation equals
\[
 \max_{s\in\{0,1\}^v}\frac1{vK}\sum_{w,j}s_wA_{wj}y_j
 \le\max_{\substack{s\in\{0,1\}^v\\z\in\{-1,1\}^K}}
 \underbrace{\frac1{vK}\sum_{w,j}s_wA_{wj}z_j}_{H_{s,z}}.
\]
The inequality holds since a linear function of $y$ attains its maximum over the cube $[-1,1]^K$ at a vertex $z\in\{-1,1\}^K$. 
The right-hand side ranges over all possible $s$ and all possible sign vectors $z$ and takes the largest value, so it dominates the expression for every $y$, whatever $y$ is and however it was chosen; in particular it no longer depends on $y$, only on $A$. Each entry $A_{wj}$ is a Rademacher variable, so $\E e^{aA_{wj}}=\cosh a\le e^{a^2/2}$. With independence of the entries,
\[
 \E e^{\lambda H_{s,z}}
 =\prod_{w,j}\E\exp\Bigl(\frac{\lambda s_wz_j}{vK}A_{wj}\Bigr)
 \le\exp\Bigl(\frac{\lambda^2\sum_{w,j}s_w^2z_j^2}{2v^2K^2}\Bigr)
 \le\exp\Bigl(\frac{\lambda^2}{2vK}\Bigr),
\]
since $\sum_{w,j}s_w^2z_j^2\le vK$. There are $2^{v+K}$ pairs $(s,z)$, so, for every $\lambda>0$,
\[
 \E\max_{s,z}H_{s,z}
 \le\frac1\lambda\log\sum_{s,z}\E e^{\lambda H_{s,z}}
 \le\frac{(v+K)\log2}{\lambda}+\frac\lambda{2vK}.
\]
Minimizing over $\lambda$ proves the claim.
\end{proof}

\textbf{The upper bound via privacy:} Fix a group $i\in[g]$ and one of its hidden records, $\xi_{ia}$ with $a\in[\kappa]$. Let $V_{ia}=V\setminus\bigl(W_i\cup\{\xi_{ib}:b\ne a\}\bigr)$, $|V_{ia}|=u-m+1$, and let $A_{ia}=(p_{ij}(x))_{x\in V_{ia},\ j\in[K]}$ be the random sign matrix of these values. Write $\Omega_{ia}$ for the conditioned information: the hidden slots $J_i$ of $X_i$, its core rows $W_i$ with their slots and core signs, the other hidden values $\xi_{ib}$, $b\ne a$, with their signs in every block, and all sampling randomness of the other groups. Given $\Omega_{ia}$ the value $\xi_{ia}$ is uniform on $V_{ia}$ (which value of $V_{ia}$ occupies the slot $a$ of $X_i$ is not given) and independent of the matrix $A_{ia}$, whose entries are independent fair signs, also independent of $\Omega_{ia}$. Given $\Omega_{ia}$ and $A_{ia}$, every sign assignment $p_{\cdot j}$ of every block is determined, hence so is the whole list of subset queries; call it $\mathcal L$. The list does not depend on the value of $\xi_{ia}$.

For a candidate $w\in V_{ia}$ let $X^{(w)}$ be the complete $N$-row database in which $w$ occupies the slot $a$ of $J_i$, everything else being as fixed by $\Omega_{ia}$, and let $Y^{(w)}=(Y^{(w)}_{ij})_{j\in[K]}$ be the decoded answers of group $i$ when the mechanism runs on $X^{(w)}$ with the list $\mathcal L$. The replacement database $X^{(0)}$ has the fixed public value $z_0$ in the same slot, with the same tag, and everything else unchanged. Let $Y^{(0)}=(Y^{(0)}_{ij})_j\in[-1,1]^K$ be the decoded answers of group $i$ in the run on $X^{(0)}$ with the same list $\mathcal L$. Once $\Omega_{ia}$ and $A_{ia}$ are fixed, everything the run on $X^{(0)}$ receives is fixed as well: the database $X^{(0)}$ and the list $\mathcal L$, neither of which involves $\xi_{ia}$. Hence $Y^{(0)}$ is a function of the mechanism's coins alone, and it is independent of $\xi_{ia}$, whose only remaining randomness is the uniform choice within $V_{ia}$; in particular it is already known to differ from every row of $X^{(0)}$ outside the replaced slot.

\emph{Step 1: privacy for a fixed candidate.} Fix $\Omega_{ia}$, $A_{ia}$ and $w$. Since $X^{(w)}$ and $X^{(0)}$ are neighboring databases and decoding is postprocessing, the laws of $Y^{(w)}$ and $Y^{(0)}$ are $(\eps,\delta)$-close. Write $c^{(0)}_i(w)=\frac1K\sum_jp_{ij}(w)Y^{(0)}_{ij}$ for the same expression computed from $Y^{(0)}$. The map $y\mapsto\bigl(\frac1K\sum_jp_{ij}(w)y_j\bigr)_+$ is the same for the two runs, since it is determined by the list $\mathcal L$, which is identical in both, and takes values in $[0,1]$, so with expectations over the mechanism's coins only,
\[
 \E_{\mathrm{coins}}\bigl[(c_i(\xi_{ia}))_+\ \big|\ \Omega_{ia},A_{ia},\ \xi_{ia}=w\bigr]
 \ \le\ e^\eps\,\E_{\mathrm{coins}}\bigl[(c^{(0)}_i(w))_+\ \big|\ \Omega_{ia},A_{ia}\bigr]+\delta .
\]

\emph{Step 2: averaging over the candidate.} Average the inequality of Step~1 over $w$ uniform on $V_{ia}$. The two sides are treated differently because they serve different purposes. On the left, the goal is to bound the expectation over a single choice of $\xi_{ia}$. The average over $w$ is therefore turned into that expectation: as $\xi_{ia}$ is uniform on $V_{ia}$ given $\Omega_{ia},A_{ia}$, the law of total expectation gives
\[
 \frac1{|V_{ia}|}\sum_{w\in V_{ia}}\E_{\mathrm{coins}}\bigl[(c_i(\xi_{ia}))_+\ \big|\ \Omega_{ia},A_{ia},\ \xi_{ia}=w\bigr]
 =\E_{\xi_{ia},\,\mathrm{coins}}\bigl[(c_i(\xi_{ia}))_+\ \big|\ \Omega_{ia},A_{ia}\bigr].
\]
On the right, the goal is to apply Lemma~\ref{lem:random-sign-positive-parts}, which bounds an average over all rows of a matrix; the average over $w$ is therefore kept and moved inside the expectation (linearity). Since all summands refer to the same run on $X^{(0)}$, and thus to the same vector $Y^{(0)}$, this gives the average, over all rows of $A_{ia}$, of the positive correlation of $Y^{(0)}$ with the row, exactly the form the lemma bounds:
\begin{align*}
 \MoveEqLeft\frac1{|V_{ia}|}\sum_{w\in V_{ia}}\Bigl(e^\eps\,\mathop{\E}\limits_{\mathrm{coins}}\bigl[(c^{(0)}_i(w))_+\ \big|\ \Omega_{ia},A_{ia}\bigr]+\delta\Bigr)\\
 &=e^\eps\,\mathop{\E}\limits_{\mathrm{coins}}\Biggl[\frac1{|V_{ia}|}\sum_{w\in V_{ia}}\Bigl(\frac1K\sum_jp_{ij}(w)\,Y^{(0)}_{ij}\Bigr)_+\ \Bigg|\ \Omega_{ia},A_{ia}\Biggr]+\delta .
\end{align*}
Note that the lemma cannot be applied to the left side directly: there, changing the candidate $w$ changes the database $X^{(w)}$ and hence the answers $Y^{(w)}$, so the summands involve a different vector for every row, whereas the lemma requires a single vector $y$ correlated with all the rows of the matrix. The run on $X^{(0)}$, which is the same for every $w$, is what provides such a single vector, but it lives in a different world from the one we need to bound; privacy is the only bridge between the two, and this is why the argument cannot do without it.

\emph{Step 3: the lemma.} Fix $\Omega_{ia}$ and the coins. Then $Y^{(0)}$ is a function of $A_{ia}$, a $|V_{ia}|\times K$ matrix of independent fair signs, and Lemma~\ref{lem:random-sign-positive-parts} with $A=A_{ia}$, $v=|V_{ia}|=u-m+1$ and $y=Y^{(0)}$ (arbitrary dependence of $y$ on $A$ is allowed) gives
\[
 \mathop{\E}\limits_{A_{ia}}\Biggl[\frac1{|V_{ia}|}\sum_{w\in V_{ia}}\Bigl(\frac1K\sum_jp_{ij}(w)\,Y^{(0)}_{ij}\Bigr)_+\ \Bigg|\ \Omega_{ia},\mathrm{coins}\Biggr]
 \ \le\ \sqrt{2\log2\Bigl(\frac1K+\frac1{u-m+1}\Bigr)} .
\]
The bound does not depend on $\Omega_{ia}$ or on the coins, so averaging over them and combining with Steps~1 and~2,
\[
 \mathop{\E}\limits_{\Omega_{ia},A_{ia},\xi_{ia},\mathrm{coins}}\bigl(c_i(\xi_{ia})\bigr)_+
 \ \le\ e^\eps\sqrt{2\log2\Bigl(\frac1K+\frac1{u-m+1}\Bigr)}+\delta ,
\]
and averaging over $(i,a)$ preserves the right-hand side. Comparing with the lower bound \eqref{eq:add-scorelower} on the same quantity and rearranging gives:
\begin{equation}
 \frac1{10m}\le
 \frac{48\sqrt g}{m}\alpha
 +\frac{e^\rho}{\sqrt{g\kappa}}
 +e^\eps\sqrt{2\log2\left(\frac1K+\frac1{u-m+1}\right)}
 +\delta+2\eta.
 \label{eq:ordinary-core-final-simplified}
\end{equation}

\subsection{Parameters and arbitrary database sizes}
\label{sec:ordinary-parameters}

\begin{proof}[Completion of the proof of Theorem~\ref{thm:querylower}]
Set $g=\lceil A_\rho m\rceil$ and $u=K=\lceil B_\eps m^2\rceil$, where, for instance, $A_\rho=32000\,e^{2\rho}$ and $B_\eps=10^4e^{2\eps}$, and let $c=1/2000$, $\alpha\le c/\sqrt g$ and $\delta,\eta\le1/(gm)$. With $\kappa=m/20$ we get $g\kappa\ge1600e^{2\rho}m^2$, so $e^\rho/\sqrt{g\kappa}\le1/(40m)$. For large $m$, $u-m+1\ge u/2$ and $K=u$, so $\frac1K+\frac1{u-m+1}\le\frac3u$ and
\[
 e^\eps\sqrt{2\log2\Bigl(\frac1K+\frac1{u-m+1}\Bigr)}
 \le e^\eps\sqrt{\frac{6\log2}u}
 \le\frac{e^\eps\sqrt{6\log2}}{\sqrt{B_\eps}\,m}
 =\frac{\sqrt{6\log2}}{100m}<\frac{0.021}m<\frac1{40m}.
\]
The accuracy term is $\frac{48\sqrt g}m\alpha\le\frac{48}{2000m}=\frac{0.024}m<\frac1{40m}$, and $\delta+2\eta\le\frac3{gm}=o(1/m)$ because $g$ grows with $m$. The right-hand side of \eqref{eq:ordinary-core-final-simplified} is therefore at most $\frac3{40m}+o(1/m)$, while the left-hand side is $\frac4{40m}$: a contradiction for all sufficiently large $m$.

Thus, with $g=\lceil A_\rho m\rceil$ we have $N_0=gm=\Theta_\rho(m^2)$, so $m=\Theta_\rho(N_0^{1/2})$ and $c/\sqrt g=\Theta_\rho(m^{-1/2})=\Theta_\rho(N_0^{-1/4})$: accuracy $c_0N_0^{-1/4}$ is impossible for a constant $c_0$ depending only on $\eps,\rho$. The tagged universe has size $gu=\Theta_\rho(m)\cdot\Theta_\eps(m^2)=\Theta_{\eps,\rho}(m^3)=\Theta_{\eps,\rho}(N_0^{3/2})$, and the number of subset queries is $KM=\Theta_\eps(m^2)\cdot O(g\log(m+2))=O_{\eps,\rho}(m^3\log m)=O_{\eps,\rho}(N_0^{3/2}\log(N_0+1))$; the theorem allows larger universes and longer lists: the hard list is extended to the required length by empty-set queries, and a lower bound for it refutes a guarantee for all lists of that length.

For an arbitrary sufficiently large $N$, choose $m$ to be the largest multiple of $40$ with $N_0=m\lceil A_\rho m\rceil\le N$; by maximality the next multiple already exceeds $N$, that is, $N<(m+40)\lceil A_\rho(m+40)\rceil$. Thus, 
\[
 \frac N{N_0}
 <\frac{m+40}{m}\cdot\frac{\lceil A_\rho(m+40)\rceil}{\lceil A_\rho m\rceil}
 \le\frac{m+40}{m}\cdot\frac{A_\rho(m+40)+1}{A_\rho m}
 =\Bigl(1+\frac{40}m\Bigr)\Bigl(1+\frac{40}m+\frac1{A_\rho m}\Bigr)
 =1+O\Bigl(\frac1m\Bigr).
\]
Thus $N/N_0\le2$ for large $m$, and in particular $N_0=\Theta(N)$.

Pad the hard database with $N-N_0$ copies of a public symbol $\star$ of the universe that no query contains; then $|S_{\rm pad}\cap Q|=|S\cap Q|$ for every query while the normalization changes from $N_0$ to $N$:
\[
 \frac{|S_{\rm pad}\cap Q|}{N}=\frac{N_0}N\cdot\frac{|S\cap Q|}{N_0}.
\]
Multiplying the mechanism's answers by $N/N_0$ and clipping to $[0,1]$ therefore turns a mechanism for $N$-row databases into one for the hard $N_0$-row database. Replacement adjacency and prefix forgetting are preserved, and an error of $\alpha$ becomes at most $(N/N_0)\alpha\le(1+O(1/m))\alpha$, a constant factor. Also $\delta,\eta\le1/N\le1/N_0$. Finally, a larger universe is handled by embedding the hard universe in it; the additional elements appear in no database row and in no query, so the argument is unchanged.
\end{proof}

\begin{remark}[A weaker forgetting condition]
\label{rem:block-forgetting}
Prefix forgetting enters the proof only through
\eqref{eq:ordinary-core-forgetting-transfer}, and only in a restricted form.
The erased prefix always consists of whole blocks, so the cut falls on the
boundary between two blocks, and the comparison concerns only the joint
answers of the block that follows, not the entire remaining suffix. The proof
therefore goes through for every mechanism satisfying this weaker requirement,
which asks for less than Definition~\ref{intro:def:prefix}. Moreover, the
erased blocks need not be replaced by the dummy $\bot$: any fixed public
queries that merely fill the erased positions will do, for instance empty-set
queries, which are ordinary subset queries.
\end{remark}

\begin{proof}[Proof of Corollary~\ref{cor:analyst-lower}]
Run the hard query list with each block issued by a different analyst, so
that the answers to block $j$ form the view of analyst $j$. In the reference
execution for block $j$, replace all preceding blocks by empty-set queries,
keeping their analyst IDs and positions. This changes only the queries of the
other analysts, so many-to-one analyst-privacy implies that the view of
analyst $j$ is $(\rho,\eta)$-close in the two executions. This is exactly the
comparison required in Remark~\ref{rem:block-forgetting}, and the rest of the
proof of Theorem~\ref{thm:querylower} applies unchanged. The construction has
$K=\Theta(N)$ blocks, so $CN$ analyst IDs suffice.\footnote{For
the set/add-remove formulation of~\cite{DNV12}, compare each hidden
record with its deletion in the privacy argument and use distinct
public padding records. Accuracy on the smaller database is not needed.}
\end{proof}

\begin{remark}[Tightness]
\label{rem:query-tightness}
For fixed $\eps>0$ and the fixed-list, per-query accuracy guarantee
used here, the error bound is tight up to constant factors when
$|U|=\Theta(N^{3/2})$.
Indeed, adding independent $\Lap(2/\eps)$ noise once to each histogram
count is $(\eps,0)$-DP under replacement, and answering every query
from this histogram is stateless. Each fixed query has normalized
error $O(\sqrt{|U|}/(\eps N))$ with probability at least $0.99$.
Thus the universe size is necessary up to constant factors.
Likewise, independent Gaussian noise answers $k$ fixed queries with
$(\eps,1/N)$-DP and per-query error
$O(\sqrt{k\log N}/(\eps N))$, so the query count is necessary up to
polylogarithmic factors.
\end{remark}

\end{document}